\documentclass[prb,twocolumn,showpacs,preprintnumbers]{revtex4}
\usepackage{graphicx}

\usepackage{graphicx}
\usepackage{subfigure}
\usepackage{amsfonts}
\usepackage{amssymb}
\usepackage{amsmath}
\begin{document}

\title{Resonant acousto-optics in the terahertz range:
TO-phonon polaritons driven by an ultrasonic wave}
\date{\today}

\author{E.\,A. Muljarov}\email[]{egor.muljarov@astro.cf.ac.uk}
\altaffiliation[]{On leave of absence from General Physics Institute RAS, Moscow, Russia.}
\author{R.\,H. Poolman}
\author{A.\,L. Ivanov}
\affiliation{School of Physics and Astronomy, Cardiff University,
The Parade, Cardiff CF24 3AA, United Kingdom}

\begin{abstract}

The resonant acousto-optic effect is studied both analytically and
numerically in the terahertz range where the transverse-optical
(TO) phonons play the role of a mediator which strongly couples
the ultrasound and light fields. A propagating acoustic wave
interacts with the TO phonons via anharmonic channels and opens
band gaps in the TO-phonon polariton energy dispersion that
results in pronounced Bragg scattering and reflection of the
incoming light. The separation in frequency of different Bragg
replicas, which is at the heart of acousto-optics, allows us to
study the resonant acousto-optic effect in the most simple and
efficient geometry of collinear propagation of electromagnetic and
ultrasonic waves. The acoustically induced energy gaps, Bragg
reflection spectra, and the spatial distribution of the electric
field and polarization are calculated for CuCl parameters, in a
wide range of frequencies and intensities of the pumping acoustic
wave. Our results show drastic changes in terahertz spectra of
semiconductor crystals that opens the way for efficient and
accessible manipulation of their infrared properties, by tuning
the parameters of the acoustic wave.

\end{abstract}


\maketitle

\section{Introduction}

Acousto-optics is a well established discipline and has a
relatively long history which started in 1922 from a pioneering
work by Leon Brillouin,\cite{Brillouin22} who predicted light
diffraction by an acoustic wave (AW) propagating in a medium. Ten
years later this prediction received clear experimental
confirmation,\cite{Debye32,Lucas32} however, the dielectric
constant modulation caused by the photoelastic effect turned out
to be extremely small (of order of $10^{-5}$). Such a modulation
leads to a similarly small light diffraction effect, which is
normally seen only in the lowest diffraction order. This picture
is typical for conventional, non-resonant
acousto-optics.\cite{Korpel97} In contrast, {\em resonant}
acousto-optics, which has been introduced very
recently,\cite{Ivanov01,Cho05,Ivanov08,Poolman10} deals with
acoustically induced spectral changes which are comparable to the
background signal. This is achieved by a resonant coupling of the
incoming light to such fundamental excitations in the solid state
as excitons or transverse optical (TO) phonons leading to the
formation of polaritons.\cite{Hopfield58} These excitations, in
turn, are strongly coupled to the AW via the deformation potential
(in case of excitons) or phonon anharmonicities (in case of TO
phonons). This results in drastic changes of the optical
properties of solid state structures when they are exposed to an
ultrasonic wave. The predicted strong resonant acousto-optic
effect has been recently observed in semiconductor microcavities
parametrically driven by a surface AW.\cite{Lima05,Lima06} A
similar effect has been proposed and observed in photonic crystal
structures, in which the distributed feedback cavity enhances the
light-sound
interaction.\cite{Santos01,deLima03,deLima04,Krishnamurthy04,deLima05,Courjal10}

A travelling AW produces a propagating periodic perturbation in a
solid state structure. To some extent, such periodicity is similar
to that in a photonic crystal. Thus the Bragg diffraction of light
typical for photonic crystal slabs is expected also in structures
with travelling modulations. In the case of an acoustically
induced grating the periodic modulation propagates in space with
the speed of sound. Therefore we are dealing with a {\it dynamical
photonic crystal} structure which is essentially different from
conventional photonic crystals. Apart from a perfect periodicity
induced by the AW, an obvious advantage of this dynamical
structure is the ability to tune the Bragg lattice constant by
changing the frequency of the pumping AW. A further advantage
stems from a fundamental property of these dynamical structures,
that the Bragg replicas, which can be spatially resolved in the
far field, also acquire a spectral resolution. In this case the
scheme for the detection of Bragg scattering can be greatly
simplified by allowing the light and the AW to propagate
collinearly, therefore, using only the spectral/temporal
resolution. We employ such a scheme in the present paper assuming
both light and ultrasound waves propagating in the same/opposite
direction that allows us to reduce the strong-coupling theory of a
three-body problem to an effective one-dimensional  dynamical
problem.

The purpose of this paper is to study in detail the far-infrared
properties of TO-phonon polaritons driven by an ultrasonic wave,
by solving (numerically) exactly a pair of Maxwell's and material
macroscopic equations. A CuCl semi-infinite crystal is taken as an
example in all simulations presented in the paper. Since the
TO-phonon resonant energy typically ranges from several to tens of
milli-electron-volts (20.3\,meV in CuCl) we concentrate in this
paper on polariton properties in the far infrared (terahertz)
range.

We show, in particular,  that owing to the anharmonic coupling to
TO phonons, a coherent ultrasonic excitation gives rise to
travelling periodic modulations in semiconductor structures and to
dynamical Bragg scattering of electromagnetic waves. In unbound
semiconductors, we describe this scattering in terms of
polaritonic partial plane waves which include all important Bragg
harmonics and calculate the quasienergy spectrum of such
excitations. The translational invariance of the quasienergy
spectrum brings into resonance different polariton states which
strongly couple to each other. As a result of such anticrossing of
different polariton branches, band gaps induced by the coherent AW
open in the quasienergy polariton spectrum and manifest themselves
as maxima in the reflection which are in turn efficiently tunable
by changing parameters of the pump AW. Expanding the
electromagnetic field and the TO-phonon polarization into partial
polaritonic waves of unbound system, we calculate exactly (with a
given accuracy) the full reflectivity spectra as well as frequency
selected particular Bragg replicas and their interference and
analyze in detail their dependence on the AW parameters.

We have also developed a perturbative analytic approach which
allows us to calculate acoustically induced band gaps and Bragg
reflection spectra and analyze them in terms of phonon-assisted
transitions between different TO-phonon polariton states.
Comparing analytic results with the full numerical calculation of
the reflection, we demonstrate that the analytical approximation
works well for low and moderate excitation power of the AW. Based
on the translational symmetry of the AW-driven polariton
quasi-energy spectrum, the analytic approach allows us to show
explicitly that the acousto-optic effect is resonantly enhanced in
down-converted Bragg replicas. The maximum enhancement occurs at
the frequencies satisfying the Bragg condition,\cite{Korpel97} due
to the polariton scattering by the AW accompanied by emission of
acoustic phonons.

The paper is organized as follows. Section II presents the general
formalism of the approach, introducing the macroscopic equations
valid for collinear propagation of the light and ultrasound waves
and their solutions in terms of plane waves. The quasi-energy
spectrum of bulk TO-phonon polaritons driven by an AW is
calculated and analyzed in Sec.\,III. The vacuum/semiconductor
interface is introduced in Sec.\,IV where the frequency-resolved
Bragg replicas and the full reflectivity spectrum are calculated,
both analytically and numerically. In Sec.\,V, we discuss in more
detail how the terahertz properties of semiconductors change with
acoustic frequency, paying attention to the limits of very high
and very low acoustic frequencies as well as to decay and
redistribution of the electric field. Results are summarized in
Sec.\,VI. Some important details of the analytic approach and
numerical methods used in the present work are given in Appendices
A\,--\,E.

\section{Formalism}
The propagating AW is excited through the piezoelectric component
of the electron-phonon interaction. At the same time, it couples
to TO-phonons via anharmonic channels. The lowest order effect is
the cubic phonon anharmonicity, which is described by the
following interaction, here written in terms of phonon
displacement operators:\cite{Cowley68,Menendez84}
\begin{equation}
V=\sum_{\bf k\,p} \Phi^{(3)}({\bf k},{\bf p})
 (b_{\bf p}+b_{\bf -p}^{\dag} ) (b_{\bf k-p}+b_{\bf -k+p}^{\dag})
 ( a_{\bf -k}+a_{\bf k}^{\dag} )\,,
  \label{V-cubic1}
\end{equation}
where $a_{\bf k}$ and $b_{\bf k}$ are the acoustic and TO-phonon
annihilation operators, respectively. In case of a propagating
ultrasonic pump wave, the acoustic-phonon operators in
Eq.~(\ref{V-cubic1}) are replaced by their classical expectation
values giving
\begin{equation}
 V = \sigma \sum_{\bf p} b_{\bf p} b_{\bf -K+p}^{\dag}e^{i\Omega
 t} +\ {\rm H.\,c.}\,,
 \label{V-cubic2}
\end{equation}
where off-resonant terms are neglected (they are however taken
into account below in the macroscopic equations). The coupling
strength
 \begin{equation}
\sigma\propto\sqrt{I_{\rm ac}}\Phi^{(3)}
 \label{sigma}
 \end{equation}
is proportional to the AW amplitude ($I_{\rm ac}$ is the acoustic
intensity) and to the Fourier transform of the third-order
derivative of the interatomic potential $\Phi^{(3)}$. A careful
investigation\cite{Poolman10} shows that, to good approximation,
any dependence on phonon wave vectors in $\Phi^{(3)}$ can be
safely neglected. The operating acoustic frequencies are all
limited to the linear regime of the acoustic phonon dispersion
given by
 \begin{equation}
\Omega(K) = v_{\rm s} K\,,
 \end{equation}
where ${\bf K}$ is the acoustic wave vector ($K=|{\bf K}|$) and
$v_{\rm s}$ is the sound velocity.

The macroscopic polarization ${\bf P}({\bf r},t)$ due to the TO
phonons is introduced as the inverse Fourier transform of the
expectation value of the TO-phonon displacement operator, $\langle
b_{\bf p}(t)+b_{-\bf p}^{\dag}(t)\rangle$. Then, using the fact
that ${\bf D}=\varepsilon_b {\bf E} + 4\pi {\bf P}$, where ${\bf
E}$ is the electric field and $\varepsilon_b$ is the background
dielectric constant, the TO-phonon polariton in the field of a
coherent pump AW is described by a system of coupled Maxwell's and
material equations.\cite{Ivanov01,Poolman10} In what follows, we
concentrate on collinear propagation of the acoustic and light
waves, normal to the semiconductor surface. In case of a
semiconductor occupying half space $z>0$, the system can be
effectively reduced to a one-dimensional geometry and is described
by the following differential equations ($\hbar=1$ is used for
simplicity of notations)
\begin{eqnarray}
 &&\left(\frac{\varepsilon_b}{c^2}
\frac{\partial^2}{\partial t^2} - \frac{\partial^2}{\partial z^2}
\right) E(z,t) = - \frac{4 \pi}{c^2}\,\frac{\partial^2}{\partial
t^2} \, P(z,t)
\label{Maxwell} \\
 &&\left[\frac{\partial^2}{\partial t^2} + 2 \gamma
 \frac{\partial}{\partial t} + \omega^2_t + 4\sigma \omega_t \cos(
Kz-\Omega t)\right]P(z,t)\nonumber\\
 &&\hskip3.7cm =\frac{\varepsilon_b\,\omega_R^2}{4 \pi}
 E(z,t),
\label{macro}
\end{eqnarray}
where $\omega_t$ and $\gamma$ are, respectively, the frequency and
the damping of bulk dispersionless TO phonons, $\omega_R$ is the
Rabi frequency which characterizes the coupling strength between
light and TO phonons. Equations (\ref{Maxwell}) and (\ref{macro})
generalize the famous polaritonic Hopfield model~\cite{Hopfield58}
by adding the interaction with the acoustic field. They can be
derived with the help of Heisenberg equations of motion from the
Hamiltonian of TO-phonons polaritons.\cite{Huang51,Romero99} An
essential difference to the conventional polariton model is the
appearance of an oscillating term in Eq.\,(\ref{macro}), which
originates from the TO-phonon coupling to the coherent AW
described by Eq.\,(\ref{V-cubic2}).

We solve Eqs.\,(\ref{Maxwell}) and (\ref{macro}) in the
semiconductor region in terms of partial plane waves. Due to the
oscillating pump acoustic field, the electric field and
polarization of each partial wave can be expanded into the
following Fourier series
 \begin{eqnarray}
E(z,t)&=&e^{i(kz-\omega t)}\sum_n E_n e^{in(Kz-\Omega t)}\,,
 \label{En}
\\
P(z,t)&=&e^{i(kz-\omega t)}\sum_n P_n e^{in(Kz-\Omega t)}\,,
 \label{Pn}
 \end{eqnarray}
where $\omega$ and $k$ are, respectively, the leading frequency
and wave vector of the partial wave of the electro-magnetic field.
The acoustically-induced perturbation is periodic both in time and
space and thus it is formally required to make a double Fourier
transform. However, as shown in Appendix~\ref{A}, this leads to
linearly dependent combinations of plane waves, so that the double
Fourier transform of $E$ and $P$ is effectively reduced to the
form of Eqs.\,(\ref{En}) and (\ref{Pn}). We substitute them into
Eqs.\,(\ref{Maxwell}) and (\ref{macro}) and equate coefficients at
the same exponentials. This results in the following matrix
eigenvalues problem:
 \begin{eqnarray}
&&\hspace{-10mm}\left[
(k+nK)^2-\frac{\varepsilon_b}{c^2}(\omega+n\Omega)^2\right]
E_n=\frac{4\pi}{c^2}\,(\omega+n\Omega)^2 P_n\,,
 \label{Maxwell-En}
 \\
&&\hspace{-10mm}\Bigl[\omega_t^2-2i\gamma
(\omega+n\Omega)-(\omega+n\Omega)^2\Bigr]
P_n\nonumber\\
&&\hspace{5mm}+2\sigma\omega_t(P_{n+1}+P_{n-1})=\frac{\varepsilon_b\,\omega_R^2}{4\pi}\,E_n\,.
 \label{macro-En}
 \end{eqnarray}
We solve the set of coupled linear equations
Eqs.\,(\ref{Maxwell-En}) and (\ref{macro-En}) and find the
polariton dispersion $\omega(k)$ and the corresponding
eigenvectors $E_n$ and $P_n$ which are the amplitudes of the
partial plane waves of the TO-phonon polariton propagating in a
semiconductor modulated by the AW.

The matrix eigenvalue problem formulated in
Eqs.\,(\ref{Maxwell-En}) and (\ref{macro-En}) is in general
complex, and the values of the light frequency $\omega$ and/or
momentum $k$ can be complex even if the damping $\gamma$ is
neglected. Here we distinguish two different physical problems:
(i) A modified polariton quasi-energy spectrum where for any given
real value of the polariton momentum $k$ we find the complex
polariton energy $\omega$, as done in Sec.\,\ref{sec-quasienergy}
below. (ii) The far infrared properties of the acoustically-driven
TO-phonon polaritons, such as reflection, scattering, and
transmission, therefore we are dealing with the real frequency
$\omega$ of the incoming light. In case (ii), considered in
Sec.\,\ref{sec-reflectivity} in detail, the wave vector $k$ is
complex-valued as it describes propagation and attenuation of the
polariton waves in the medium.

In all numerical simulations done in the present work we use the
following parameters of bulk CuCl:\cite{Wakaki07} TO-phonon energy
$\omega_t=20.28\,$meV, Rabi splitting $\omega_R=14.53\,$meV, low
temperature phonon damping\cite{Ulrich99,Cardona01}
$\gamma=0.2$\,meV (at $T=5-10$\,K), background dielectric constant
$\varepsilon_b=5.2$, and sound velocity $v_s=2.02\cdot 10^3$\,m/s.
The coupling constant $\sigma$, one of the important parameters of
the model, scales with the acoustic intensity according to
Eq.\,(\ref{sigma}), with the value of $\sigma=1$\,meV
corresponding\cite{Poolman10} to the acoustic intensity $I_{\rm
ac}=14.8$\,kW/cm$^2$.

\section{Quasi-energy spectrum of acoustically driven TO-phonon polaritons}
 \label{sec-quasienergy}

\subsection{Eigenfrequencies}
 \label{sec-eigenfrequencies}

In unbound semiconductors, the polariton quasi-momentum $k$ is a
good quantum number which takes real values. For nonzero damping
$\gamma$, the polariton eigenfrequencies, in turn, are always
complex. Let us however concentrate on propagating solutions and
disregard, for the moment, energetic band gaps where propagation
is not allowed. Then, assuming $\gamma=0$, all eigenfrequencies
also become real. This slightly idealized picture of undamped
polaritons in unbound semiconductors will help us, however, to
understand the basic properties of the polariton dispersion
modified by the AW.

Treating $k$ as a real parameter, we solve the eigenvalue problem,
Eqs.\,(\ref{Maxwell-En}) and (\ref{macro-En}), finding polariton
eigenfrequencies $\omega=\tilde{\omega}_\lambda(k)$ which
correspond to the energy bands $\lambda=1,2,\dots$ separated by
acoustically-induced band gaps. Technically, we truncate
Eqs.\,(\ref{Maxwell-En}) and (\ref{macro-En}) at $-n_{\max}
\leqslant n \leqslant n_{\max}$ introducing the matrix dimension
$M=2n_{\max}+1$. The system of two $M\!\times\! M$ matrix
equations, containing nonlinear dependence both on frequency and
wave vector is reduced, in Appendix~\ref{B}, to the following {\em
linear} matrix eigenvalue problem
 \begin{equation}
\hat{\mathbb W}\vec{\mathbb X} =\tilde{\omega} \vec{\mathbb X}\,,
 \end{equation}
solved by means of complex matrix diagonalization, where
$\hat{\mathbb W}$ is an $\tilde{\omega}$-independent
$4M\!\times\!4M$ hypermatrix and $\vec{\mathbb X}$ is a
$4M$-hypervector containing the Fourier components of the electric
field and polarization, defined in Eqs.\,(\ref{AppB-matrix}) and
(\ref{AppB-vector}), respectively. Note that the extra factor of 4
in the matrix dimension $4M$ comes from two different sources. The
existence of two branches in the bare polariton spectrum, the
so-called upper polariton (UP) and lower polariton (LP) branches
provide a factor of 2. The fact that all eigenfrequencies come in
pairs\cite{footnote1} ($\tilde{\omega}_\lambda$ and
$-\tilde{\omega}^\ast_\lambda$) provides the remaining factor of
2.

Clearly, the quasi-energy spectrum $\tilde{\omega}_\lambda(k)$ has
the following translational property valid {\it within the same
branch} $\lambda$:
 \begin{equation}
\tilde{\omega}_\lambda(k+sK)=\tilde{\omega}_\lambda(k)+s\Omega\,,
 \label{transl-a}
 \end{equation}
where $s$ is an integer. The corresponding eigenfunctions are also
invariant in a way that, for the same $\lambda$,
 \begin{equation}
E_{n}(k+sK)=E_{n+s}(k)\,,
 \label{transl-b}
 \end{equation}
and a similar expression is valid for $P_n$.

\begin{figure}[t]
\includegraphics[angle=0,width=0.99\linewidth]{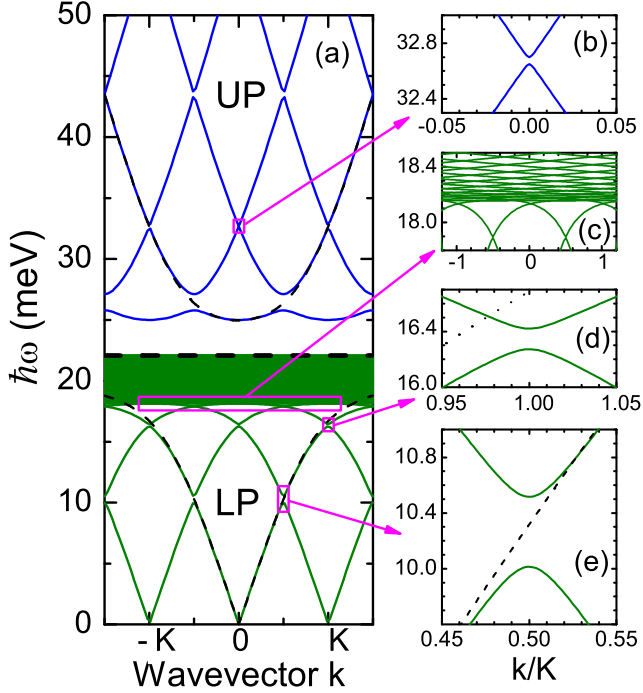}
\caption{Quasi-energy spectrum of UP (blue) and LP (green)
branches of undamped TO-phonon bulk polaritons driven by an AW
with $\nu=\Omega/(2\pi)= 100$\,MHz ($K=0.3109\,\mu$m$^{-1}$) and
$\sigma = 1$\,meV. The spectrum without AW, $\omega=\omega^0(k)$,
is given by thin dashed curves. Thick dashed horizontal line shows
an artificial cut-off in the quasi-continuum of LP states. Panels
(b)\,--\,(e) highlight particular areas in the spectrum. }
\end{figure}

The translationally invariant polariton spectrum
$\tilde{\omega}_\lambda(k)$ is shown in Fig.\,1 as a function of
the polariton momentum $k$ (measured in units of the acoustic wave
vector $K$). Due to the spatial and temporal periodicity imposed
by the pumping AW, the bare polariton spectrum (dashed curves) is
folded into the first Brillouin zone and then translational
shifted, both horizontally and vertically, into all other
Brillouin zones. The coupling of polaritons to the AW opens gaps
in the folded polariton spectrum, just in the center and on the
borders of the Brillouin zone, see Fig.\,1\,(b), (d), and (e).
These gaps are acoustically induced spectral regions where
polariton propagation is not allowed and thus they crucially
affect the terahertz properties of the polaritons, such as light
reflection and Bragg scattering. The positions of the gaps are
determined mainly by the frequency $\Omega$ (or wave vector $K$)
of the pumping AW whereas their widths strongly depend on both the
wave vector $K$ and the intensity (coupling strength $\sigma$) of
the AW, as discussed in detail in the next section.

In the polariton bands, which are the regions where polariton
propagation is allowed, the modified spectrum deviates from the
bare one. This is illustrated in Figs.\,1\,(d) and (e). This
deviation become dramatic close to and above the TO-phonon
resonant energy. Unlike the bare polariton dispersion in which the
LP branch is bounded by the TO-phonon frequency, the AW-induced
branches of the quasi-energy spectrum, even those which originate
from the LP branch, can be found at any frequencies. Thus, they
are in principle unbound. They form a sort of a quasi-continuum of
polariton states detailed in Fig.\,1\,(c), which is due to the
acoustically-induced anticrossing and level repulsion. The optical
strength of these quasi-continuum states, however, drops quickly
with energy and contributes to the reflectivity only at the
frequencies around the low-energy edge of the Restrahlen band.

\subsection{Acoustically-induced band gaps}
 \label{sec-gaps}

The band gaps are spectral regions forbidden for polariton
propagation, therefore one can anticipate an enhancement of
reflection at the frequencies of incoming light close to or inside
the gaps. The analytic calculation of acoustically induced band
gaps, which we show in this section, is particularly important for
the understanding of this enhancement and other terahertz
properties of acoustically driven polaritons, considered later in
Sec.\,\ref{sec-reflectivity}.

The tridiagonal form of the matrix in Eq.\,(\ref{macro-En}) allows
us to evaluate the widths of the energy gaps analytically.
Expressing $P_n$ from the diagonal Eq.\,(\ref{Maxwell-En}) and
substituting it into Eq.\,(\ref{macro-En}) results in
 \begin{equation}
\alpha_n D_n E_n-\sigma
\Bigl(\alpha_{n+1}E_{n+1}+\alpha_{n-1}E_{n-1}\Bigr) = 0
 \label{tri}
 \end{equation}
with
 \begin{eqnarray}
D_n(k,\omega)&=&\frac{1}{2\omega_t}\biggl[\omega_t^2-2i\gamma(\omega+n\Omega)-
(\omega+n\Omega)^2
 \nonumber\\
 && -\left.
\frac{\omega_R^2}{\alpha_n(k,\omega)}\right],
 \label{Dn}
\\
\alpha_n(k,\omega)&=&
\frac{c^2}{\varepsilon_b}\left(\frac{k+nK}{\omega+n\Omega}\right)^2-1\,.
 \label{alpha}
 \end{eqnarray}
Equation (\ref{tri}) can be solved recursively. Introducing
$\xi_n=E_{n}/E_{n-1}$, it is transformed to
 \begin{equation}
\xi_n=\frac{\sigma \alpha_{n-1}}{\alpha_n
D_n-\sigma\alpha_{n+1}\xi_{n+1}}\,.
 \label{recursive}
 \end{equation}
With a truncation $E_{-N-1}=E_{1}=0$, Eq.\,(\ref{tri}) is solved
up to the $N$-th order. The ratio $\sigma/D_n\sim
2\sigma/\omega_t$ serves as a smallness parameter in our
perturbation approach which breaks down when the frequency
$\omega$ approaches either the TO-phonon resonance $\omega_t$ or
the bottom of the UP branch, $\sqrt{\omega_t^2+\omega_R^2}$.

The above truncation is equivalent the following boundary
conditions added to the recursive Equation~(\ref{recursive}):
 \begin{equation}
\xi_{-N-1}=1/\xi_{0}=0\,.
 \label{BCs-xi}
 \end{equation}
Taking the first boundary condition \mbox{$\xi_{-N-1}=0$} as a
staring point, we find $\xi_{-N}$, $\xi_{-N+1}$, and so on, ending
up with the relation $\alpha_{0}D_{0}=\sigma\alpha_{-1}\xi_{-1}$,
in which the second boundary condition is used. The above
procedure turns out to be particularly useful for a perturbative
approach to the resonant acousto-optic effect discussed in
Sec.\,\ref{sec-refl-analyt}, where the Bragg replicas are
calculated analytically in lowest orders. In the present section
we evaluate perturbatively only the acoustically induced band
gaps, again assuming $\gamma=0$.

\begin{figure}[t]
\includegraphics[angle=0,width=0.95\linewidth]{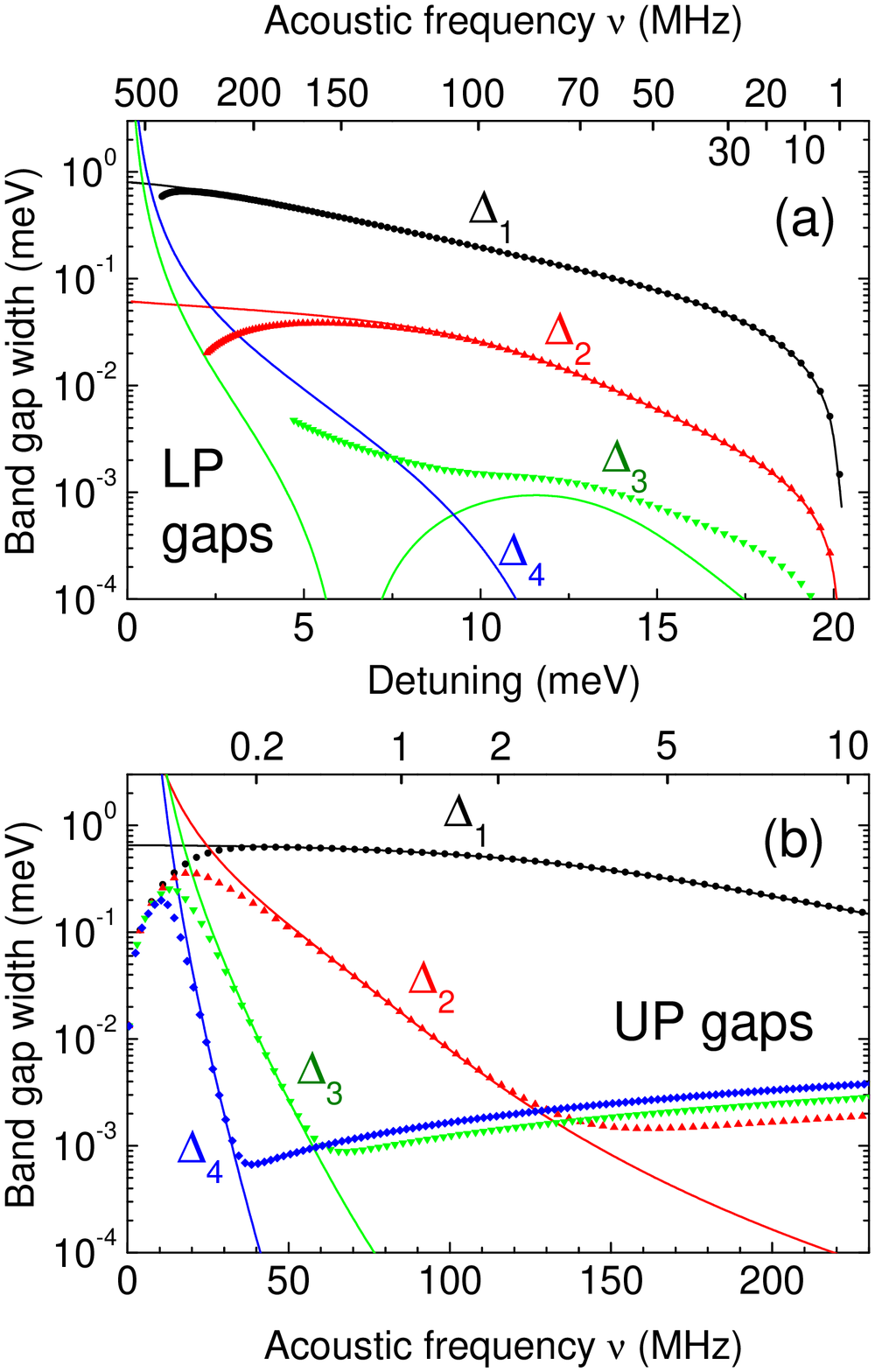}
\caption{(a) Analytical (curves) and numerical (symbols)
calculation of the acoustically induced band gaps in the LP branch
vs the acoustic frequency $\nu=\Omega/(2\pi)$ (top axis) and the
detuning $\omega_t-\omega_1$ (bottom axis). (b) Energy gaps in the
UP branch vs acoustic frequency (bottom axis) and the detuning
$\omega_1-\sqrt{\omega_t^2+\omega_R^2}$ (top axis). In both cases
(a) and  (b), $\omega_1$ is the energy position of the lowest gap,
either in the LP (a) or in the UP part (b) of the spectrum. The
coupling strength $\sigma=0.4$\,meV, the TO-phonon damping
$\gamma=0$.}
\end{figure}

In the $N=0$ case, Eqs.\,(\ref{recursive}) and (\ref{BCs-xi})
result in a single equation
 \begin{equation}
D_0(k,\omega) = 0\,,
 \label{zeroth}
 \end{equation}
which yields the bare polariton dispersion $\omega=\omega^0(k)$
consisting of the LP and UP branches (see dashed curves in
Fig.\,1), with no effect from the AW. The acoustically induces
gaps appear in the polariton spectrum in first and higher
perturbation orders at specific values of the polariton wave
vector given (neglecting the acoustic frequency $\Omega$) by the
famous Bragg condition
 \begin{equation}
k_N=NK/2\,,\ \ \ N=1,\,2\,\dots,
 \label{Bragg-condition}
 \end{equation}
which corresponds to centers and borders of different Brillouin
zones (obviously all of them can be translated into the first
Brillouin zone). The energy positions of the gaps, $\omega_{N}$,
are given (in zeroth order) by the bare polariton dispersion,
 \begin{equation}
\omega_{N}=\omega^0(k_N)\,,
 \label{gap}
 \end{equation}
or, in terms of Eq.\,(\ref{zeroth}), can be found by solving the
equation $D_0(k_N,\omega_{N}) = 0$. The widths of the gaps,
$\Delta_{N}$, are evaluated in lowest order in Appendix~\ref{C},
with the following result:
 \begin{eqnarray}
\Delta_1&=&\sigma S(\omega_1)\,,
\nonumber\\
\Delta_2&=&\sigma^2\,\frac{ S(\omega_2)}{|D_{-1}(k_2,\omega_2)|}\,,\nonumber\\
\Delta_3&=&\sigma^3\,\frac{S(\omega_3)}{D^2_{-1}(k_{3},\omega_3)}\,,\nonumber\\
\Delta_4&=&\sigma^4\,\frac{S(\omega_4)}{D^2_{-1}(k_{4},\omega_4)|D_{-2}(k_{4},\omega_4)|}\,,\nonumber\\
&&\dots \label{gapN}
 \end{eqnarray}
where
 \begin{equation}
S(\omega)=\frac{2\omega\omega_t\omega_R^2}{(\omega^2-\omega_t^2)^2+\omega_t^2\omega_R^2}\,.
 \end{equation}

The gap $\Delta_N$ opens in the $N$-th perturbation order as a
result of anticrossing of different polariton dispersion branches.
The anticrossings occur due to the translational invariance of the
spectrum which brings into resonance states originating from
negative-$k$ and positive-$k$ parts of the bare polariton
dispersion, e.g. by a translational shift in the
($k,\omega$)-plane of the states in the negative-$k$ part of the
dispersion by the vector ($NK$,$N\Omega$), i.e. by the same
integer of the wave vector and the frequency of the AW. In other
words, the $N$-th gap opens due to the $N$-acoustic phonon
transition, induced by the pumping AW, which couples two point in
the bare dispersion $\omega^0(k)$ having the opposite values of
$k$ as given by the Bragg condition Eq.\,(\ref{Bragg-condition}).
As a result, the width of the gap $\Delta_N$ is proportional to
$\sigma^N$ (and to $I_{\rm ac}^{N/2}$). The formfactor $S(\omega)$
grows by increasing the TO-phonon component and reducing the light
component of the polariton, hence the gaps become wider. In fact,
since TO phonons play the role of a mediator between the light and
acoustic fields (in our model, these two fields do not interact
directly), and thus the more the component of TO-phonons in the
polariton field, the stronger the resonant acousto-optic effect.

Figure 2 presents a comparison of the analytical and numerical
calculations of the band gaps at a moderate acoustic intensity
$I_{\rm ac}=2.4$\,kW/cm$^2$ ($\sigma=0.4$\,meV). The analytic
results for the widths of the two lowest band gaps, $\Delta_1$ and
$\Delta_2$ (lines), are in excellent agreement with the full
numerical calculation (symbols), practically in the whole range of
the acoustic frequencies and both in the UP and LP parts of the
spectrum, as seen from Fig.\,2. At smaller detuning which
corresponds to higher (lower) acoustic frequency in the LP (UP)
branch, the agreement gets worse because the analytical model,
which is based on the assumption that the detuning is large
compared to the coupling strength, breaks down in this case. At
the same time, higher-order band gaps like $\Delta_3$ and
$\Delta_4$ are not well reproduced by the analytics. This can be
explained by the fact that the present analytic approach neglects
the acoustic frequency $\Omega$, assuming that it is small
compared both to the light frequency and to the energy gaps (for
details, see Appendix~\ref{C}). This, however, is not the case of
the third- and fourth-order band gap, as their widths fall into
sub-$\mu$eV-range and thus become comparable to or even smaller
than $\Omega$.

\section{Bragg reflection of electro-magnetic waves}
 \label{sec-reflectivity}

\subsection{Polariton quasi-momenta}
 \label{sec-quasimomentum}

The quasi-energy spectrum and the band gaps discussed in detail in
Sec.\,\ref{sec-quasienergy} can be probed in experiments by a
monochromatic electromagnetic wave with frequency $\omega$. In
this case, the boundary between the semiconductor and the vacuum
becomes an essential part of the system, so that the momentum
across the surface is no longer conserved. However, for the given
frequency $\omega$ of incoming light, the polariton waves can have
only specific values of the complex momentum,
$k=\tilde{\kappa}_j(\omega)$. They are known in the literature as
forced harmonic solutions.\cite{Tait72} The complex quasi-momenta
$\tilde{\kappa}_j(\omega)$ can be found as eigenvalues of the same
Equations~(\ref{Maxwell-En}) and (\ref{macro-En}). In this case we
solve the inverse problem to Sec.\,\ref{sec-eigenfrequencies}: the
complex quasi-momenta $\tilde{\kappa}_j(\omega)$ are found as
eigenvalues. The electric field and polarization of the
propagating polariton waves are the corresponding normalized
eigenvectors, $E_{nj}$ and $P_{nj}$, where an extra index
$j=0,\,\pm1,\,\pm2,\dots$ is added in order to distinguish
different partial waves. For each frequency, Maxwell's boundary
conditions determine a unique combination of the polariton partial
waves and finally the reflection and transmission properties which
will be discussed in subsequent sections. The partial waves
themselves are however not fully independent. For any integer
number $s$, the wave vectors and amplitudes of the partial waves
obey the following translational relations [cf. with
Eqs.(\ref{transl-a}) and (\ref{transl-b})]
 \begin{eqnarray}
\tilde{\kappa}_{j + s}(\omega) &=& \tilde{\kappa}_j(\omega - s
\Omega) + sK\,,
 \label{k-transl}
\\
E_{n,j+s}(\omega)&=&E_{n+s,j}(\omega - s \Omega)\,,
 \label{E-transl}
\\
P_{n,j+s}(\omega)&=&P_{n+s,j}(\omega - s \Omega)
 \label{P-transl}
  \end{eqnarray}
following from the spatio-temporal periodicity of the system
pumped by an AW, periodic in time and space.

\begin{figure}[t]
\includegraphics[angle=0,width=0.95\linewidth]{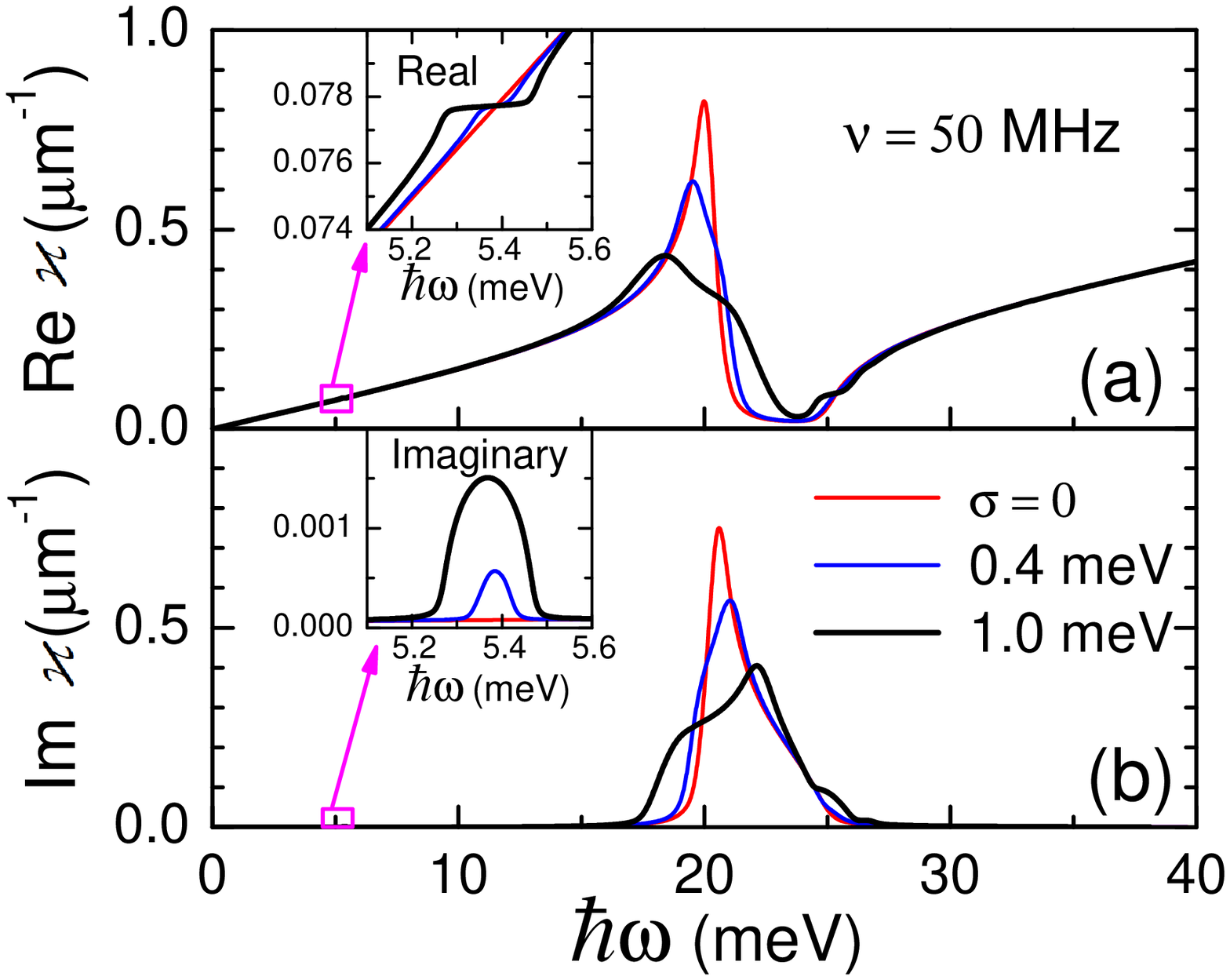}
\caption{(a) Real part and (b) imaginary part of the polariton
quasi-momentum $\varkappa(\omega)$ calculated with [black and blue
(dark gray) curves] and without AW [red (light gray) curves]. The
insets highlight particular regions corresponding  to the lowest
acoustically induced gap. The TO-phonon damping $\gamma=0.2$\,meV
and the AW frequency $\nu=50$\,MHz ($K=0.1555\,\mu$m$^{-1}$).}
\end{figure}

The quasi-momenta $\tilde{\kappa}_j$ are calculated in the
following way.  The infinite-matrix problem
Eqs.\,(\ref{Maxwell-En}) and (\ref{macro-En}) is again truncated
to $-n_{\max} \leqslant n \leqslant n_{\max}$ ($M=2n_{\max}+1$).
The amplitude $P_n$ is expressed from Eq.\,(\ref{Maxwell-En}) and
substituted into Eq.\,(\ref{macro-En}). The latter equation is
then linearized with respect to $\tilde{\kappa}$, as done in
Appendix~\ref{B}, and brought to the form
 \begin{equation}
\hat{\mathbb V}\vec{\mathbb Y} =\tilde{\kappa} \vec{\mathbb Y},
 \end{equation}
with a $2M\!\times\!2M$ hypermatrix $\hat{\mathbb V}$ and a
$2M$-hypervector $\vec{\mathbb Y}$ defined in Eqs.\,(\ref{appB-V})
and (\ref{appB-Y}), respectively. There is now only one extra
factor of 2 in the matrix dimensions $2M$. This is due to the
symmetry $\tilde{\kappa}\to-\tilde{\kappa}$ of the initial
eigenvalue problem. However, there is no more clear separation of
the eigenvalues into LP and UP branches.

\begin{figure}[t]
\includegraphics[angle=0,width=0.99\linewidth]{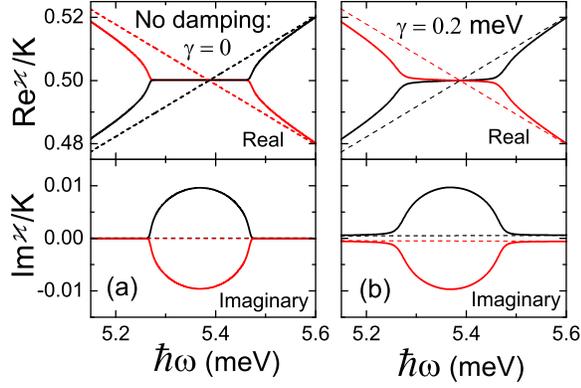}
\caption{Real and imaginary parts of the polariton quasi-momentum
$\varkappa(\omega)$ calculated around the acoustically induced gap
(a) for zero damping $\gamma=0$ and (b) for finite damping
$\gamma=0.2$\,meV, with (solid lines) and without AW (dashed
lines). Black and red (gray) colors correspond to positive-$k$
unshifted and negative-$k$ one TA-phonon shifted polariton
dispersion branches, respectively. The AW frequency
$\nu=50$\,MHz.}
\end{figure}

Technically, we calculate the eigenvalues $\tilde{\kappa}_j$ for
quite big matrices (with $M=161$, so that $-80\leqslant j
\leqslant 80$), for different values of $\omega$ and then select
only the central root
 \begin{equation}
\varkappa(\omega)\equiv \tilde{\kappa}_{j=0}(\omega)\,,
 \label{kappa}
 \end{equation}
i.e. the root which satisfies the equation
\mbox{$\varkappa(0)=0$}. The central root is chosen as best
calculated. In fact, it is the most distant solution from the
boundaries of the truncated array, so that any distortions of the
spectrum due to such truncation are minimized in the center of the
array. We then use this central root $\varkappa$ to improve
calculation of all other wave vectors $\tilde{\kappa}_j$ ($j\neq
0)$ exploiting the translational property Eq.\,(\ref{k-transl})
and making a linear interpolation Eq.\,(\ref{interp}) between
neighboring frequency points.

Both the real and imaginary parts of the central root $\varkappa$
are shown in Fig.\,3, in the presence of AW and without it. Apart
from very strong changes in the polariton momentum slightly below
the TO-phonon resonant frequency (20.3\,meV) and in the whole
Restrahlen band above it, tiny changes occur near the acoustically
induced gaps and grow with increasing coupling strength $\sigma$,
see the insets to Fig.\,3. These small features, however, are very
important for the infrared properties of the system and thus we
consider them in more detail.

The real part of the quasi-momentum [Fig.\,4(a), top panel], if it
were calculated for $\nu=100$\,MHz, would reproduce the
quasi-energy spectrum shown in Fig.\,1(e) for zero damping. The
only difference would be the horizontal connection line in
Fig.\,4(a). However, the results shown in Figs.\,3 and 4 are
calculated for half that acoustic frequency, which has moved the
gap to energies nearly half that in Fig.\,1(e). The energy gap is
due to the acoustically induced strong coupling between the two
bare polariton branches crossed in the center of Fig.4(a), see
dashed lines. The real part of the momentum shows the typical
anticrossing behavior while the imaginary part, strictly absent at
$\gamma=\sigma=0$, demonstrates a pronounced band gap. Polariton
waves for which the frequencies fall in the gap region propagate
only into a limited area due to their finite decay in space, which
is determined by the imaginary part of the wave vector. With
finite damping (which is still smaller than the band gap width),
the situation does not change much, but the imaginary part of the
wave vector remains finite everywhere in this case, see Fig.4(b).

\subsection{Bragg replicas: Numerical approach}
 \label{sec-refl}

Let us now consider the effects of reflection, transmission and
Bragg scattering due to a semiconductor/vacuum boundary at $z=0$.
In the semiconductor area ($z>0$), the electric field and
TO-phonon polarization are superpositions of partial polaritonic
waves,
 \begin{eqnarray}
E(z,t)&=&\sum_j A_j E_j(z,t)\nonumber \\
&=&\sum_{nj} A_j E_{nj}
e^{i(\tilde{\kappa}_j+nK)z-i(\omega+n\Omega t)}\,,
 \label{En2}
\\
P(z,t)&=&\sum_{nj} A_j P_{nj}
e^{i(\tilde{\kappa}_j+nK)z-i(\omega+n\Omega t)}\,,
 \label{Pn2}
 \end{eqnarray}
where $E_{nj}$ and $P_{nj}$ are normalized eigenvectors and
$\tilde{\kappa}_j$ are eigenvalues described and calculated in
Sec.\,\ref{sec-quasimomentum}. In the vacuum region ($z<0$), the
electric field is given by
\begin{equation}
E(z,t) = e^{iq_0z} e^{-i\omega t} + \sum_{n}{r_n e^{-iq_nz}
e^{-i(\omega + n \Omega)t}} \, ,
 \label{vacuum}
\end{equation}
where $q_{n}=(\omega + n \Omega)/c$ is the wave vector and $r_n$
is the amplitude of the outgoing Bragg replica $n$. The amplitude
$r_n$ is normalized to the unity amplitude of the incoming light
wave, which is described by the first term in the right hand side
of Eq.\,(\ref{vacuum}). Coefficients $A_j$, which stand for the
amplitudes of the partial waves, are found from Maxwell's boundary
conditions. These require continuity of the electric and magnetic
fields, $E(z,t)$ and $H(z,t)$, across the interface $z=0$.
Applying the boundary conditions to the electric field
Eq.\,(\ref{En2}) and (\ref{vacuum}) and to the magnetic field
calculated via $\partial H/\partial t=c\,\partial E/\partial z$
results in the following set of equations
 \begin{eqnarray}
&&e^{-i\omega t}+\sum_n r_n e^{-i(\omega+n\Omega)t} =\sum_{nj} A_j
E_{nj} e^{-i(\omega+n\Omega)t}\,,\nonumber\\
&&e^{-i\omega t}-\sum_n r_n
e^{-i(\omega+n\Omega)t}\nonumber\\
&&\hspace{25mm}
 =\sum_{nj} A_j E_{nj}
\frac{\tilde{\kappa}_j + nK}{q_n}
e^{-i(\omega+n\Omega)t}\,,\nonumber
 \end{eqnarray}
which have to be satisfied at any time $t$. Thus we equate
coefficients of the same exponentials, which then leads to a set
of linear algebraic equations for $r_n$ and $A_j$:
\begin{eqnarray}
\delta_{n,0}+r_n &=& \sum_j A_j E_{nj}\,,
 \label{bc1}
\\
\delta_{n,0} - r_n &=& \sum_j A_j E_{nj}\frac{\tilde{\kappa}_j +
nK}{q_n}\,.
 \label{bc2}
\end{eqnarray}
\begin{figure*}
\includegraphics[angle=0,width=0.99\linewidth]{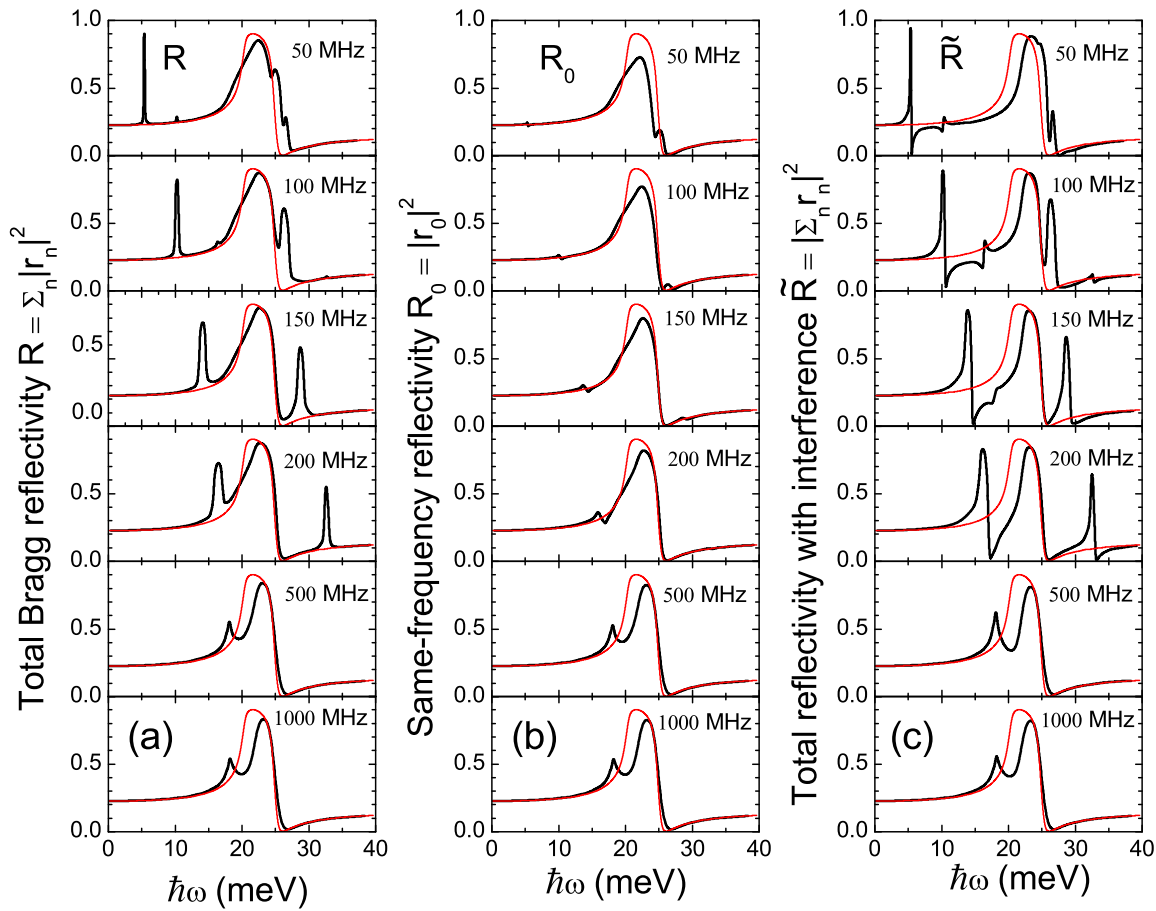}
\caption{(a) Incoherent and (c) coherent total Bragg reflectivity
calculated for $\sigma=1$\,meV and different acoustic frequencies,
without and with interference terms, respectively. (b) The same
frequency central Bragg replica. The bare reflection spectrum of
damped TO-phonons (with no AW) is shown by red (gray) thin
curves.}
\end{figure*}
Equations (\ref{bc1}) and (\ref{bc2}) are solved numerically,
using a truncation $-n_{B} \leqslant n \leqslant n_{B}$ with the
total number of Bragg replicas $N_B=2n_B+1$. Normally, the number
of Bragg replicas $N_B$ required for an accurate calculation of
the reflectivity spectra is much smaller than the matrix dimension
$M$ used in the calculation of the polariton quasi-momentum
$\varkappa(\omega)$, $N_B\ll M$. For example, to reach the
accuracy $10^{-7}$ in calculation of CuCl reflectivity using the
AW parameters $\sigma = 2\,$meV and $\nu=100$\,MHz, one needs to
take only $N_B=21$ and $M=81$ into account.

\subsection{Reflectivity spectra}
 \label{sec-refl-numer}

In our geometry, the light and the AW propagate collinearly and
thus there is no way to have any spatial selection of the Bragg
replicas. However, each Bragg replica has it own unique temporal
evolution as can be seen in Eq.\,(\ref{vacuum}). Thus, using a
frequency selective technique one can measure the intensity of the
$n$-th Bragg replica in the reflected light,
 \begin{equation}
R_n=|r_n|^2\,.
 \end{equation}
The $n$-th replica carries the frequency $\omega+n\Omega$, it is
down-converted (up-converted) with respect to the frequency
$\omega$ of the incoming light, for negative (positive)~$n$.

\begin{figure*}
\includegraphics[angle=0,width=0.67\linewidth]{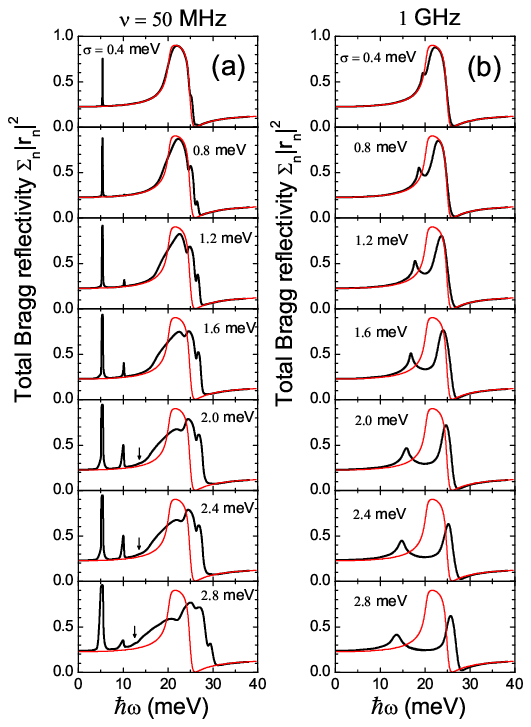}
\caption{Total Bragg reflectivity $R(\omega)$ calculated for the
acoustic frequency $\nu=\Omega/(2\pi)=50$\,MHz (a) and 1\,GHz (b),
and for different values of the coupling strength $\sigma$
(acoustic intensity $I_{\rm ac}$). Small vertical arrows in panel
(a) point at the spike growing due the third-order band gap. The
bare reflection spectrum of damped TO-phonons is shown by red
(gray) thin curves.}
\end{figure*}

Without such frequency selection, the total but still spectrally
resolved Bragg reflection can be measured. Due to the different
time evolution of different Bragg signals, there can be two
opposite ways to measure the reflectivity. One is fully
incoherent, in this case the temporal resolution $\tau$ of the
spectrometer is insufficient to see the interference fringes
between different Bragg replicas. In other words, the interference
terms fully vanish in the measured electric field. This happens
when $\Omega\tau\gg 1$, so that $\tau$ falls into a nanosecond
range. In this case, the total Bragg reflectivity is given by
 \begin{equation}
R=\sum_{n=-n_B}^{n_B} |r_n|^2.
 \end{equation}
Alternatively, the spectrometer can have enough temporal
resolution and then the opposite limit $\Omega\tau\ll 1$ is
realized making the total Bragg reflectivity fully coherent:
 \begin{equation}
\widetilde{R}=\left|\sum_{n=-n_B}^{n_B} r_n\right|^2.
 \end{equation}
Such regime is achievable when the spectrometer resolution is in
the sub-nanosecond or even picosecond range.

We show in Fig.\,5 both the coherent and incoherent signals,
$\widetilde{R}(\omega)$ and $R(\omega)$, in comparison with the
zeroth replica $R_0(\omega)$. Note that $R_0$ is the only
component which survives in the absence of any acoustic pumping;
for $\sigma=0$ it is shown in Fig.\,5 by a red (gray) curve, the
same in all panels. All other spectra shown in Fig.\,5 by black
curves, are calculated for $\sigma=1$\,meV. They reveal drastic
changes compared to the bare ($\sigma=0$) spectrum. By increasing
the acoustic frequency from 50\,MHz to 1\,GHz, a strong spike in
the reflectivity $R$, which is associated with the acoustically
modified LP branch moves towards the Restrahlen band, gets
thicker, and finally saturates into a cusp-like feature close to
the bare TO-phonon transition energy. Well below the TO-phonon
energy, the spike is exclusively due to the first-order
acoustically induced band gap discussed in Sec.\,\ref{sec-gaps}.
Its thickness is given either by the band gap width $\Delta_1$
[see Eq.\,(\ref{gapN})] or by the phonon damping, whichever is
greater. A similar spike which comes from the UP branch is well
resolved above the Restrahlen band, for the acoustic frequencies
between 100 and 300\,MHz. The coherent Bragg reflectivity
$\widetilde{R}$ has similar properties, however the spectral
changes are enhanced considerably by pronounced dispersive
features which appear due to the strong interference effect.

At higher acoustic frequencies, the contribution of the central
Bragg replica, $R_0$, becomes dominant and all other replicas
vanish as $\nu\to\infty$, so that all three spectra $R(\omega)$,
$R_0(\omega)$, and $\widetilde{R}(\omega)$ finally coincide. This
happens because the light wavelength becomes large compared to the
acoustic wavelength and thus the AW modulation of the
semiconductor structure is not resolved by the light field. The
acoustically modulated semiconductor works in this case as an
effective homogeneous medium with a modified $\omega$-dependent
dielectric constant. The cusp-like structure in the reflectivity
which evolves from the dispersive (N-shaped) feature in $R_0$ [see
Fig.\,5(b)] can be understood as a manifestation of the band gaps
accumulating around the TO-phonon frequency. Indeed, according to
Eqs.\,(\ref{Bragg-condition}) and (\ref{gap}), at large values of
the acoustic wave vector $K$, there must be a plenty of band gaps
stacked together around $\omega_t$.

\begin{figure*}
\includegraphics[angle=0,width=0.99\linewidth]{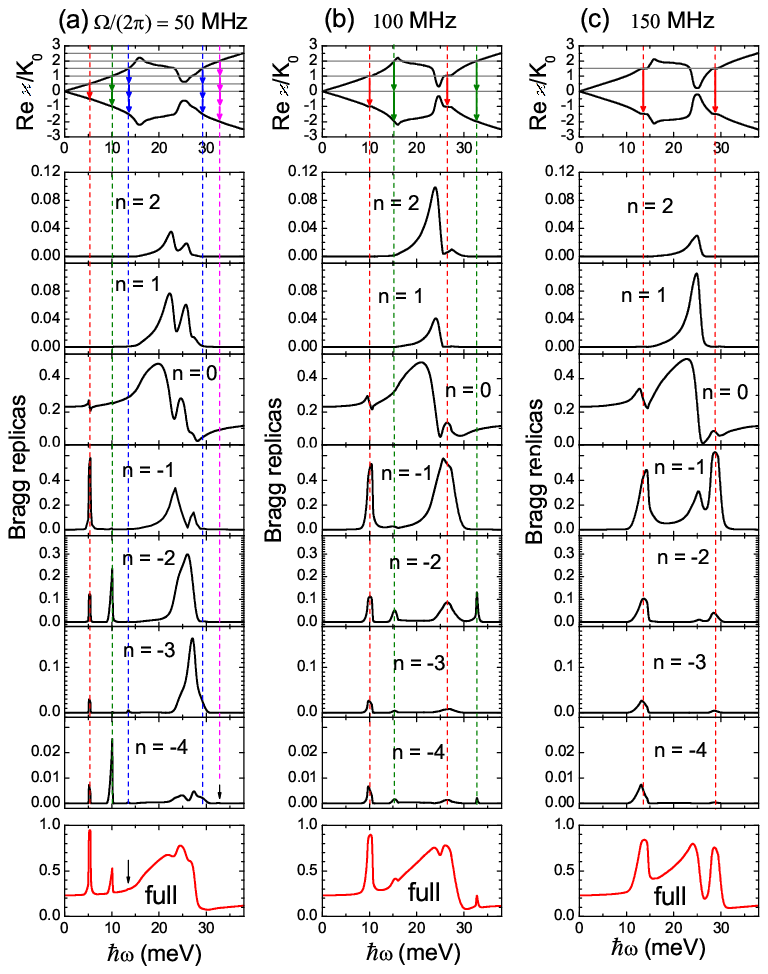}
\caption{Acoustically modified polariton dispersion of damped
TO-phonons (top panels), contributions of different Bragg replicas
$|r_n|^2$ with $n=2$, 1, 0, -1, -2, -3, and -4, as well as the
total reflectivity spectrum $R(\omega)=\sum_n|r_n|^2$ (bottom
panels), calculated for $\sigma=2$\,meV and
$\nu=\Omega/(2\pi)=50$\,MHz (a), 100\,MHz (b), and 150\,MHz (c).
Arrows and vertical lines indicate positions of the band gaps due
to $N$-phonon dressed transitions. The acoustic wave vector takes
the values of (a) $K=K_0$, (b) $K=3K_0/2$, and (c) $K=2K_0$, where
$K_0=0.1555\,\mu$m$^{-1}$. }
\end{figure*}

In this high-frequency limit, there are no qualitative changes in
the reflectivity when the coupling to the AW increases, as it is
clear from Fig.\,6(b), where the reflectivity $R$ ($R\approx R_0$)
is shown for different values of the coupling strength $\sigma$.
Indeed, in this case only the strength of the cusp, its splitting
from the Restrahlen band, and the transparency of the latter grow
with $\sigma$. In the opposite case of small acoustic frequencies,
the increasing coupling strength does lead to qualitatively new
features in the spectrum. In particular, new peaks corresponding
to the second and third acoustically induced band gaps, both in
the UP and in the LP branches, grow up in $\nu=50$\,MHz spectra as
shown in Fig.\,6(a).

The influence of the first- and higher-order band gaps on
different Bragg replicas is analyzed in more detail in Fig.\,7.
The top panels display the polariton quasi-momentum dispersion,
similar to those shown in Figs.\,3 and 4. The Bragg condition is
illustrated in Fig.\,7 by gray horizontal lines (separated by
$K/2$) and by arrows all having the same length equal to $K$. The
main spikes in the full reflectivity spectrum $R(\omega)$, which
are shown in the bottom plots, appear exactly at the energies of
the first-order band gaps. These spikes are due to one-phonon
transitions between positive-$\varkappa$ and negative-$\varkappa$
parts of the polariton dispersion $\varkappa(\omega)$. This
effect, which is already discussed above in terms of the bare
polariton dispersion $\omega^0(k)$ (see Figs.\,1 and 4), is now
demonstrated taking into account the realistic acoustically
modified dispersion $\varkappa(\omega)$. With increasing acoustic
frequency $\Omega$ and, consequently, increasing acoustic wave
vector $K$, the same Bragg condition can only be satisfied at a
higher frequency. Thus, the band gaps and the spikes are
blue-shifted, at the same time getting wider due to a larger
TO-phonon component. This is visualized by two and tree times
longer arrows in Figs.\,7(b) and (c). Below the Restrahlen band,
these arrows point to one-phonon transitions that occur,
respectively, at nearly two and three times larger frequency. The
band gaps are more pronounced at larger $\Omega$ and manifest
themselves as longer and flatter plateaus in the dispersion [see
Fig.\,7(c), top panel]. From the practical viewpoint, the effects
demonstrated in Figs.\,5--7 are nothing else than a very efficient
way of changing the terahertz properties of semiconductor
structures simply by tuning the frequency and the intensity of the
pumping AW.

The $N$-phonon transitions (with $N=2$, 3, and 4) are shown in
Fig.\,7 by a series of $N$ arrows of different color. They are
responsible for higher-order acoustically induced band gaps and
for new spikes, which appear in the reflectivity when $\sigma$
increases (see also Fig.\,6). In particular, Fig.\,7(a)
demonstrates that the spikes due to second-, third-, and
fourth-order gaps show up in the Bragg replicas $n=-2$, $-3$, and
$-4$, respectively. The higher-order Bragg replicas, however,
accumulate the contribution of all lower-$N$ phonon transitions in
higher orders. The higher-order effects of lower-$N$ gaps are
sometimes even stronger than the contribution of the same $N=|n|$
gap to the Bragg replica $r_n$, see e.g. $r_{-3}$ and $r_{-4}$
replicas in Fig.\,7 at the frequencies $\omega\approx \omega_1$
and $\omega_2$. This effect will be discussed in more detail in
Sec.\,\ref{sec-refl-analyt}.

The incoming light beam excites polaritons with positive values of
the quasi-momentum $\varkappa>0$, and then these polaritons are
scattered by the AW to the negative-$\varkappa$ part of the
polariton dispersion. Such scattering is accompanied by emission
of acoustic phonons and consequently by down-conversion in
frequency. These down-converted polariton waves correspond to
negative-$n$ components of the electric field, both in
semiconductor and in vacuum, see Eqs.\,(\ref{En2}) and
(\ref{vacuum}). That is why all spikes in the reflection
corresponding to the AW-induced band gaps, appear only in Bragg
replicas having negative $n$ and no such features can be seen when
$n$ is positive.

By increasing the coupling strength $\sigma$, $N$-phonon
transitions responsible for the band gaps reflectivity spikes
become increasingly dressed by higher order transitions. In
particular, the $N$-th order gap is formed due to multiple
processes of simultaneous emission of $N+s$ phonons and absorption
of $s$ phonons, where $s$ can be any integer. In the present
calculation however, with the parameters used throughout the
paper, this number never exceeds in reality $|s|=20$, even for
very large acoustic intensities (up to $\sigma=3$\,meV).

\subsection{Bragg replicas: Analytic approach}
 \label{sec-refl-analyt}

At low acoustic intensities (small $\sigma$), the Bragg replicas
can be calculated using the perturbative approach introduced in
Sec.\,\ref{sec-gaps}. In this approach, the $N$-th perturbation
order describes the acousto-optic effect with participation of $N$
phonons (no higher-order dressing is included), and in particular
the opening of the gaps $\Delta_N$ due to $N$-phonon transitions.
Below we calculate acousto-optic effect in each $N$-th Bragg
replica in its lowest ($N$-th) order, calculating both
up-converted $r_{+N}$ and down-converted $r_{-N}$ replicas, and
show explicitly that the effect is resonantly enhanced only in the
down-converted replicas $r_{-1}$, $r_{-2}$, etc.

The perturbation-theory result for the $N$-th Bragg replicas has
the following recursive form (for derivation see
Appendix~\ref{D}):
 \begin{eqnarray}
r_{\pm N}&=&-\sum_{s=1}^{N} A_{\pm(N-s)} E_{\pm s}\frac{\beta_{\pm
s}-\beta_0}{1+\beta_0}\,,
 \label{rpmN}
 \\
A_{\pm N}&=&-\sum_{s=1}^{N} A_{\pm(N-s)} E_{\pm
s}\frac{1+\beta_{\pm s}}{1+\beta_0}\,,
 \label{ApmN}
 \\
E_{\pm N}&=&\sigma^N\frac{\alpha_0}{\alpha_{\pm
N}}\prod_{s=1}^N\frac{1}{D_{\pm s}}\,.
 \label{EpmN}
 \end{eqnarray}
where $D_n(\omega,k)$ are defined in Eq.\,(\ref{Dn}),
 \begin{equation}
\alpha_n=\left(\frac{k+nK}{p}\right)^2-1\,,\ \ \ \
\beta_n=\frac{k+nK}{q}\,,
 \label{an}
 \end{equation}
(we have neglected here any dependence on $\Omega$), $q=\omega/c$,
and $p=\sqrt{\varepsilon_b}\,q$. The starting $N=0$ values in the
recursion are given by
 \begin{equation}
r_0=\frac{1-\beta_0}{1+\beta_0}\,,\ \ \
A_0=\frac{2}{1+\beta_0}\,,\ \ \ E_0=1\,.
 \label{r0}
 \end{equation}

To a first approximation, the quasi-momentum $k$ in the above
equations is taken in zeroth order, i.e. is calculated according
to the bare polariton dispersion
 \begin{equation}
k=k^0(\omega)\,,
 \end{equation}
which satisfies Eq.\,(\ref{zeroth}) and is inverse to
$\omega=\omega^0(k)$. Using this bare dispersion relationship,
$D_0(\omega,k)=0$, the functions $D_n(\omega,k)$ are simplified to
 \begin{equation}
  \alpha_n D_n=\frac{\omega_R^2}{2\omega_t}\,
  \frac{(k+nK)^2-k^2}{k^2-p^2}
 \label{anDn}
 \end{equation}
where any dependence on $\Omega$ is neglected.

The explicit form of Eq.\,(\ref{anDn}) allows us to analyze the
resonances seen in Bragg replicas. Since $D_n$ stands in the
denominator of the electric field components Eq.\,(\ref{EpmN}),
$r_n$ have maxima when $\alpha_n D_n$ is minimized. This happens
when $(k+nK)^2\approx k^2$, leading to a Lorentzian spectral line
shape peaked at the Bragg condition,
 \begin{equation}
    {\rm Re}(k) =-\frac{nK}{2}\,,
 \label{Bragg2}
 \end{equation}
and having the widths proportional to Im$(k)$ which at this point
is fully determined by the TO-phonon damping $\gamma$. The
incoming light excites polaritons with positive wave vectors
Re$(k)>0$, thus the Bragg condition is fulfilled only for negative
$n<0$, so that the peaks in the reflectivity are observed in the
negative (down-converted) Bragg replicas only.

The amplitudes of the three lowest-order Bragg replicas have the
following explicit form
\begin{widetext}
 \begin{eqnarray}
 r_{-1}&=&\sigma\,\frac{K}{q}\,\frac{2}{(1+\beta_0)^2}\,\frac{\alpha_0}{\alpha_{-1}D_{-1}}\,,
  \label{rm1}
  \\
 r_{-2}&=&\sigma^2\,\frac{K}{q}\,\frac{2}{(1+\beta_0)^2}\left[\frac{2\alpha_0}{\alpha_{-2}D_{-2}D_{-1}}
-\frac{1+\beta_{-1}}{1+\beta_0}\,\frac{\alpha_0^2}{\alpha_{-1}^2D_{-1}^2}\right],
  \label{rm2}
  \\
 r_{-3}&=&\sigma^3\,\frac{K}{q}\,\frac{2}{(1+\beta_0)^2}\left[\frac{3\alpha_0}{\alpha_{-3}D_{-3}D_{-2}D_{-1}}
-\frac{3+2\beta_{-1}+\beta_{-2}}{1+\beta_0}\,\frac{\alpha_0^2}{\alpha_{-2}\alpha_{-1}D_{-2}D_{-1}}
+\left(\frac{1+\beta_{-1}}{1+\beta_0}\right)^2\,\frac{\alpha_0^3}{\alpha_{-1}^3D_{-1}^3}
 \right].
  \label{rm3}
 \end{eqnarray}
\end{widetext}
The analytic result Eqs.\,(\ref{rm1})--(\ref{rm3}) is demonstrated
in Fig.\,8(a) in comparison with the full numerical calculation;
both are done for a moderate coupling strength $\sigma=0.4$\,meV.
The agreement is good everywhere except the spikes, in which only
positions are reproduced well, see a zoomed area in the inset of
Fig.\,8(a). Indeed, with purely Lorentzian peaks in the spectra,
the magnitudes of the replicas sometimes exceed unity which
results in fully unphysical reflectivity. This artefact is however
removed below and the spectrum is refined by taking into account
the acoustically modified polariton dispersion. For positive
values of $n$, the replicas are shown in Fig.\,8(b) in order to
demonstrate that no spikes at the energies of AW-induced band gaps
are seen either in analytical or in numerical calculation.

\begin{figure}[t]
\includegraphics[angle=0,width=1.1\linewidth]{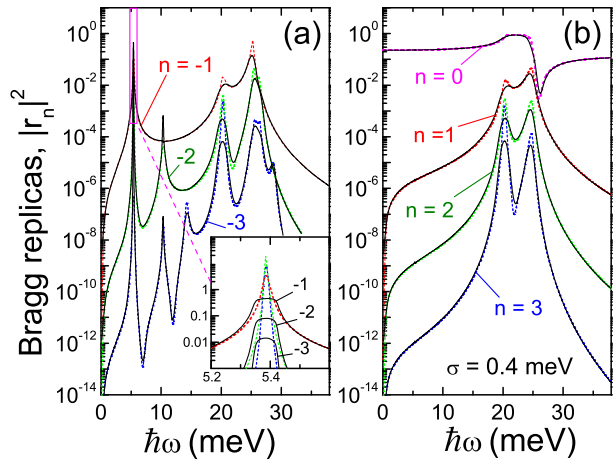}
\caption{Bragg replicas $|r_n|^2$ calculated numerically (black
lines) and analytically [red, green, blue, and magenta (gray)
lines] for (a) $n=-1$, $-2$, $-3$, and (b) $n=1$, 2, 3, and 0.
Parameters used in both calculations: the coupling strength
$\sigma=0.4$\,meV, the acoustic frequency $\nu=50$\,MHz, the
TO-phonon damping $\gamma=0.2$\,meV. The inset zooms in on the
spectral region of the first acoustically-induced band gap and
spikes in the Bragg reflection due to one-phonon transitions. }
\end{figure}

The first down-converted Bragg replica $r_{-1}$ has a spike at the
frequency $\omega_1$ of the lowest acoustically-induced energy
gap, see Eq.\,(\ref{gap}). The spike corresponds to a one-phonon
transition between polariton states in the positive-$k$ and
negative-$k$ branches of the dispersion, as shown in Fig.\,7 and
discussed earlier in Sec.\,\ref{sec-refl-numer}. Note that
although we concentrate here mainly on the LP part of the
spectrum, similar spikes can be seen also in the spectral region
of the UP branch.

\begin{figure}[t]
\includegraphics[angle=0,width=0.99\linewidth]{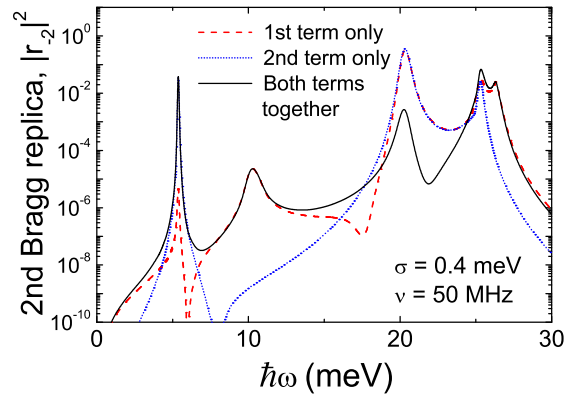}
\caption{The second Bragg replica $|r_{-2}|^2$ calculated
analytically according to Eq.\,(\ref{rm2}): the first term only
(red dashed line), the second term only [blue (gray) solid line],
and both terms together (black solid line). All parameters are the
same as in Fig.\,8. }
\end{figure}

In the second replica $r_{-2}$, resonances appear at the energies
of both first and second band gaps, $\omega_1$ and $\omega_2$. The
first term in square brackets in Eq.\,(\ref{rm2}) describes a
composite two-phonon process with a virtual intermediate state
having only one phonon emitted. That is why the first term in
Eq.\,(\ref{rm2})  shown in Fig.\,9 by a dashed red line is peaked
at both energies $\omega_1$ and $\omega_2$. The final state has
the dominant role and so the resonance at $\omega\approx\omega_2$
is much stronger. The second term shown by blue (gray) line in
Fig.\,9 has only one peak at $\omega\approx\omega_1$ (in the LP
part of the spectrum) as it describes a one-phonon contribution in
second order. This one-phonon process is the only component of the
first-order Bragg replica $r_{-1}$. Surprisingly, its dominant
role is well seen in $r_{-2}$ as well, see the sum of both terms,
the black solid line in Fig.\,9. This physical picture helps us to
understand similar features in higher
Bragg replicas. In particular, 
the first term in Eq.\,(\ref{rm3}) is due to a three-phonon
transition having as intermediate (virtual) states those with one
and two phonons emitted. The last term is the one-phonon process
taken into account in third order, and the middle term is a mixed
contribution of both one and two phonon-assisted transitions.
Consequently, the resonances in $r_{-3}$ occur at
$\omega\approx\omega_1$, $\omega_2$, and $\omega_3$ and are due to
different processes with participation of three acoustic phonons.

Let us now refine the analytic calculation in order to get rid of
the artificially high analytic reflectivity which is detailed in
the inset to Fig.\,8(a). To do this, we simply replace the bare
polariton dispersion $k=k^0(\omega)$ in Eqs.\,(\ref{an}) and
(\ref{anDn}) by the acoustically modified dispersion
$k=\varkappa(\omega)$ calculated in Sec.\,\ref{sec-quasimomentum}
[see Eq.\,(\ref{kappa})]. Consequently, a modified Bragg condition
which follows from Eq.\,(\ref{anDn}) and determines the
acousto-optic resonant energies $\omega_n$ becomes
 \begin{equation}
    {\rm Re}\bigl[\varkappa(\omega_{|n|})\bigr] =-\frac{nK}{2}
 \label{Bragg3}
 \end{equation}
[cf. with Eqs.\,(\ref{Bragg2}) and (\ref{gap})].  The refined
Bragg condition Eq.\,(\ref{Bragg3}) has been used already in our
analysis of the spectra shown in Fig.\,7. From the viewpoint of
the perturbation theory developed in the present section, the
calculation of the modified dispersion $\varkappa(\omega)$ can
intrinsically take into account a large number of higher
perturbation orders. Thus, it is not fully consistent with the
calculation of $r_n$ limited to the lowest order $|n|$ only. From
the practical viewpoint, however, such a refinement works
perfectly well for sufficiently narrow resonances, as can be seen
in Fig.\,10(a) in which the spectral lines are shown by the solid
red and green (gray) curves in comparison with full calculation
(black curves). Both the height and the width of the spikes are
reproduced well. The narrow and unrealistically high Lorentzian
lines shown by the dashed curves are fully determined by the
damping $\gamma$. In the refined analytic calculation, the
linewidth is determined by the band gap width, which is properly
incorporated into the modified dispersion $\varkappa(\omega)$. A
slight difference between the numerical and refined analytical
calculations is due to the fact that for such a moderate acoustic
intensity ($\sigma=0.4$\,meV), higher-order corrections (with
$s>|n|$) to the Bragg replicas $r_{n}$ which not included in
Eqs.\,(\ref{rpmN})--(\ref{EpmN}) are already perceptible. They
become crucial in case of much wider gaps in the UP part of the
spectrum, see Fig.\,10(b), where the improvement due to the
dispersion refinement is rather minor.

\begin{figure}[t]
\includegraphics[angle=0,width=0.99\linewidth]{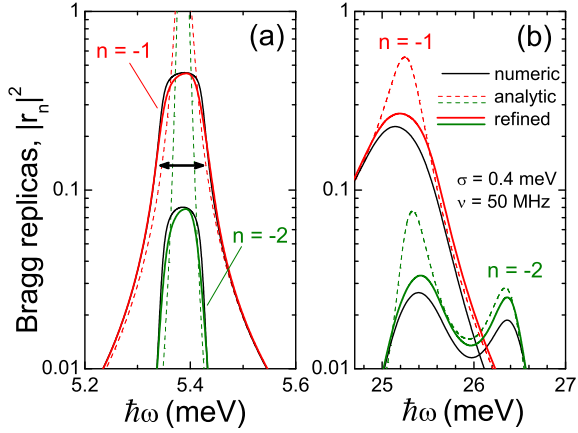}
\caption{Bragg replicas $|r_{-1}|^2$ and $|r_{-2}|^2$ calculated
analytically with [red and green (gray) solid lines] and without
(dashed lines) refinement of the polariton dispersion. The full
numerical calculation is show by black solid lines. All parameters
are the same as in Fig.\,8. The double-side arrow shows the width
of the first acoustically induced gap calculated via
Eq.\,(\ref{gapN}) }
\end{figure}

Finally, a calculation of the $n=0$ Bragg replica, which has been
done up to now only in zeroth order, can be further improved by
inclusion of second-order corrections:
 \begin{equation}
r_0=\frac{1-\beta_0}{1+\beta_0}+\sigma\left(\frac{K}{q}\right)^2\frac{4}{(1+\beta_0)^3}\,
\frac{\alpha_0^2}{\alpha_1 D_1\alpha_{-1}D_{-1}}\,.
 \label{r0improved}
 \end{equation}
 Clearly, this expression contains resonant features due to
the $D_{-1}$ in the second term. However, unlike the higher order
replicas which exhibit the Lorentzian line shape, this resonant
term is added to the background reflection, Eq.\,(\ref{r0}). Thus,
$|r_0|^2$ has instead a more pronounced dispersive behavior, due
to the mixing of the first and second terms in
Eq.\,(\ref{r0improved}). Such dispersive features manifest
themselves as N-shaped resonances clearly seen in Fig.\,5(b) and
become more and more prominent with increasing acoustic frequency
(see also $n=0$ replicas in Fig.\,7).

\begin{figure}[b]
\vskip-1cm
\includegraphics[angle=0,width=1.00\linewidth]{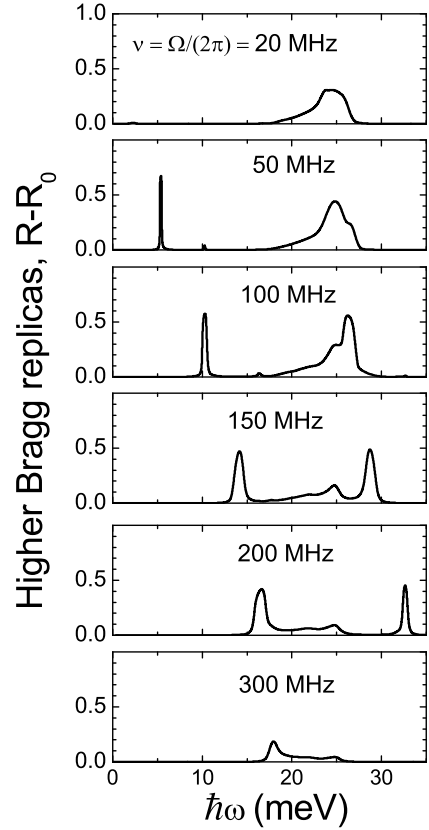}
\vskip-1cm\caption{The total contribution of higher Bragg replicas,
$\sum_{n\neq0} |r_n|^2$, for the coupling strength $\sigma=1$\,meV
and different acoustic frequencies.}
\end{figure}

\section{Discussion}
 \label{sec-discussions}

\subsection{Low and high acoustic frequency limits}

The terahertz properties of semiconductors and their changes due
to varying acoustic frequency $\Omega$ has been already discussed
in Sec.\,\ref{sec-refl-numer}. In particular, Fig.\,5(a) has
demonstrated an efficient manipulation of the spectral properties,
changing both the positions of the spikes and their widths simply
by tuning $\Omega$. Moreover, in the limiting case of very large
$\Omega$, the wavelength of the spatial modulation caused by the
AW turns out to be much smaller than the wavelength of the
propagating light, so that the modulation does not produce any
diffraction. As a result, all acoustically induced spectral
changes are concentrated in the central Bragg replica $R_0$ [cf.
Figs.\,5(a) and (b)],  while the contribution of all higher
replicas vanishes at $\Omega\to \infty$. This is more clearly
demonstrated in Fig.\,11 where the lower panels shows diminishing
contribution of the higher ($|n|>0$) Bragg replicas to the total
reflection. The results are summarized in Fig.\,12 where the
spectral area of the higher replicas is plotted against $\Omega$.
This spectral area is calculated according to
 \begin{equation}
\int_{-\infty}^\infty  \sum_{n\neq0} |r_n(\omega)|^2\,d\omega\,.
 \label{sp-area}
 \end{equation}

\begin{figure}[t]
\includegraphics[angle=0,width=0.9\linewidth]{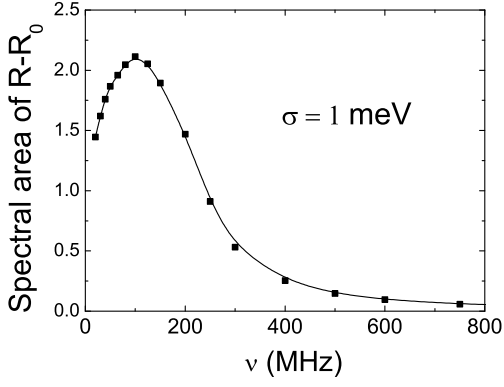}
\caption{The spectral area of higher Bragg replicas, calculated
according to Eq.\,(\ref{sp-area}) for $\sigma=1$\,meV and
different values of the acoustic frequency $\nu=\Omega/(2\pi)$. }
\end{figure}

Surprisingly, a similar effect takes place in the opposite limit
of low acoustic frequencies. Although the reduction of the
contribution of higher replicas in this case is not very well
illustrated by Fig.\,11, and technically this limit is hard to
achieve with the present numerical method, an overall trend is
such that the spectral area is getting smaller and possibly
vanishes at $\Omega\to0$, as seen in Fig.\,12. In this case, the
short-period acoustic modulations work again as a homogeneous
dielectric medium, which can be be characterized by an effective
(modified) dielectric constant. In a sense, this picture is
similar to the effective mass approximation for electrons, valid
in bulk semiconductors as well as in short-period semiconductor
superlattices where the Bragg scattering can also be neglected.

A clear bump just above the Restrahlen band seen in the top
spectrum in Fig.\,11 which is calculated for a relatively low
acoustic frequency is due to accumulation of band gaps near the
bottom of the UP branch (a similar accumulation of gaps takes
place at $\Omega\to\infty$ near the top of the LP branch; it has
been already discussed in Sec.\,\ref{sec-refl-numer}). This narrow
spectral region remains infrared-active in higher Bragg orders due
to existence of polaritons which have momenta the same order of
magnitude as the acoustic wave vector $K$ (very small in this
case) and thus can be scattered by such an AW.

\subsection{Electric field distribution and interaction length}

When the wavelength of the incident light and the period of the
acoustic modulation are comparable as shown in Fig.\,13, it
results in significant spectral changes which are mainly due to a
strong Bragg scattering of the polariton waves from the
propagating periodic patterns induced by the AW. The electric
field distribution in Fig.\,13 corresponds to the terahertz
spectrum shown in the top panel of Fig.\,5(a). While in the vacuum
region the modulations of the electric field are determined
exclusively
\begin{figure}[t]
\vskip2.5cm\hskip-5.5cm
\includegraphics[angle=270,width=0.85\linewidth]{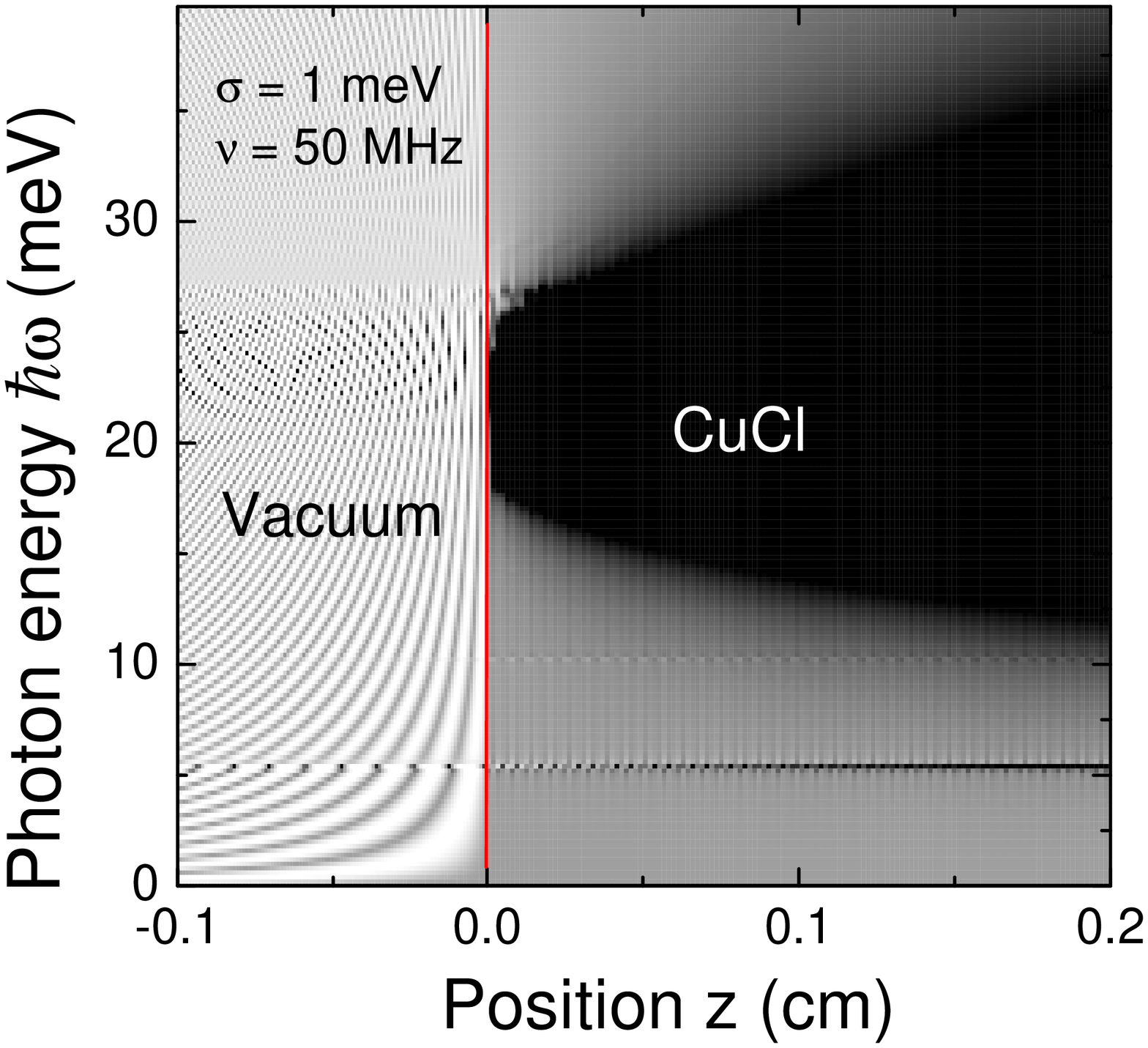}
\vskip-2.5cm
\caption{The electric field intensity $|E(z,\omega)|^2$ in vacuum
and semiconductor regions, calculated for $\sigma = 1$\,meV, $\nu
= 50$\,MHz, and different frequencies of the incoming terahertz
light. Black in the logarithmic gray scale map corresponds to
$|E|^2\to 0$. }
\end{figure}
by the light wavelength (resp. $\omega$) and are caused by a
superposition of the incoming and outgoing waves. In the
semiconductor region the period of modulations is the same for all
$\omega$ and is given by the acoustic wavelength $2\pi/K$. At
around $\hbar\omega = 5$\,meV the AW produces a half-$\lambda$
modulation ($\lambda$ is the wavelength of the light propagating
inside semiconductor) that results in a destructive interference
of co- and contra-propagating polariton (light) waves, and thus in
quick attenuation of the electric field seen as a narrow black
stripe in Fig.\,13.

\begin{figure}[t]
\includegraphics[angle=0,width=0.95\linewidth]{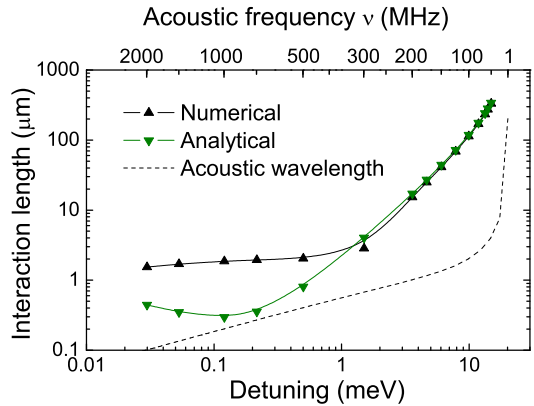}
\caption{Interaction length, corresponding to the first
acoustically-induced LP band gap $\omega=\omega_1$, calculated
analytically via Eq.\,(\ref{int-length}) and evaluated numerically
from the electric field profile, for $\sigma=1$\,meV and different
values of the acoustic frequency (top axis) or the detuning
$\omega_t-\omega_1$ (bottom axis). The dashed line shows the
acoustic wavelength.}
\end{figure}

The decay length of the electric field gives us an upper bound for
an estimate of the acousto-optic interaction length $l_{\rm int}$.
In resonant acousto-optics, such a length characterizes the scale
in real space where the light effectively couples to the
ultrasonic wave by means of the TO-phonons. The interaction length
calculated at the resonant frequency $\omega=\omega_1$ of the
first AW-induced band gap is evaluated analytically in
Appendix~\ref{E}, assuming $Kl_{\rm int}\gg 1$. Taking into account
the finite TO-phonon damping $\gamma$, it takes the following form
 \begin{equation}
l_{\rm int}=
\frac{Kc^2}{\sigma\varepsilon_b\omega_t}\,\frac{(\omega_t^2-\omega^2)^2+4\gamma^2\omega^2}{4\omega^2\omega_R^2}\,.
 \label{int-length}
 \end{equation}
Figure~14 shows the interaction length numerically evaluated from
the decay of the electric field at $\omega=\omega_1$, in the LP
part of the spectrum. It also shows an excellent agreement with
the analytic result Eq.\,(\ref{int-length}), over the whole range
of acoustic frequencies $\Omega$, or detuning
$\delta\omega=\omega_t-\omega_1$. When decreasing the detuning,
the interaction length also decreases monotonously and at very
small values of $\delta\omega$ is limited by the TO-phonon
damping. It exceeds the acoustic wavelength $2\pi/K$ (also shown
in Fig.\,14 by a dashed line) by a few tens to hundred times only.

\section{Conclusions}

We have studied, both numerically and analytically, the
far-infrared properties of acoustically pumped semiconductor
crystals, in a geometry of collinear propagation of the light and
ultrasound waves. Owing to the cubic phonon anharmonicity, TO
phonons mediate the coupling between the light and acoustic fields
that results in drastic changes of the terahertz properties of
semiconductors and in enhancement of the acousto-optic effect by
several orders of magnitude compared to the conventional
acousto-optics.

We have developed an efficient numerical method for calculation of
acoustically induced band gaps, electric field distribution, and
such infrared properties of semiconductors as transmission,
reflection, and Bragg scattering of light. The method is based on
the plane waves expansion of the electric field and TO
phonon-induced polarization. It allows us to solved the coupled
Maxwell and material equations (including a phonon anharmonic
source term due to a pumping AW) by mapping them into a linear
matrix problem. We have provided a detailed analysis of the Bragg
replicas of the incoming light (the Bragg replicas are
frequency-selected in the present case of collinear geometry) and
demonstrated a one-to-one correspondence between the spikes in the
reflectivity and the acoustically induced gaps in the polariton
quasi-energy spectrum. The gaps open at the energies $\omega_N$
satisfying the Bragg condition Eq.\,(\ref{Bragg3}), manifest
themselves as spikes in down-converted Bragg replicas $r_{n}$ with
$n=-1,\,-2,\,\dots,\,-N$, and correspond to $N$-phonon transitions
(dressed with higher orders) between different polariton states in
the crystal. In particular, the reflectivity is enhanced
considerably when the acoustic wavelength is close to half integer
($N/2$) wavelength of the light in crystal. In this case a
polariton wave is scattered by the periodic pattern of acoustic
modulation of the TO-phonon field emitting $N$ acoustic phonons.
The co-propagating polariton wave and contra-propagating scattered
wave interfere destructively to produce the electric field decay
into the crystal. We have calculated the length of such decay
which, in turn, limits the interaction length between the light
and ultrasound waves. The latter can be as short as only tens of
the acoustic wavelengths, much shorter than in conventional
acousto-optics.

Our analytic approach is based on the perturbation theory
developed in the limit of weak coupling $\sigma/\omega_t\ll 1$ and
allows us to evaluate and analyze the band gap widths, the Bragg
replicas, and the interaction length. Moreover, being in good
quantitative agreement with numerical results, the analytic
approach remains a reliable and accessible tool for calculation of
the terahertz properties up to a moderate coupling strength
($\sigma=1$\,meV in CuCl, corresponding to the acoustic intensity
$I_{\rm ac}=14.8$\,kW/cm$^2$). We have also provided a refinement
of our analytical method which uses an acoustically modified
polariton dispersion and considerably improves the calculation of
the sharp spikes in the reflectivity.

Finally, in the limits of very low and very high acoustic
frequency $\Omega$, the contribution of all higher Bragg replicas
$r_n$ with $n\neq0$ vanishes, but the central replica $r_0$ still
shows considerable spectral changes which grow dramatically with
the acoustic intensity (or with $\sigma$). In these limiting
cases, the wavelength of the propagating electro-magnetics waves
is either too small ($\Omega\to0$) or too large
($\Omega\to\infty$) compared to the acoustic wavelength to feel
the periodical modulation of the crystal, therefore, no Bragg
scattering occurs. Instead, the polariton waves propagate like in
a homogeneous medium which can be characterized by an
$\omega$-dispersive effective dielectric constant.

\acknowledgments
 We thank S.\,G. Tikhodeev and R. Zimmermann for valuable
discussions. This work was supported by the Royal Society (Grant
JP0766306), EPSRC and WIMCS.

\appendix

\section{Fourier transform of the electric field and polarization}
\label{A}

In this section, we show that the partial waves of the electric
field and TO-phonon polarization, which are found as
eigenfunctions of Eqs.\,(\ref{Maxwell}) and (\ref{macro}), can be
written in the form of a single Fourier transform,
Eqs.\,(\ref{En}) and (\ref{Pn}). Due to the periodicity introduced
by the AW, each partial wave can be expanded into Fourier series,
both in time and space. In problems similar to the quasi-energy
spectrum calculation where the wave vector $k$ is a fixed
real-valued parameter, it is convenient to use the following
space-time Fourier transform of a partial wave:
 \begin{eqnarray}
E(z,t)&=&\sum_{nm} {\cal E}_{nm} e^{i(k+nK)z}
e^{-i(\omega+n\Omega+m\Omega)t}\,,
 \label{Enm}
\\
P(z,t)&=&\sum_{nm} {\cal P}_{nm} e^{i(k+nK)z}
e^{-i(\omega+n\Omega+m\Omega)t}\,.
 \label{Pnm}
 \end{eqnarray}
Substituting the above expansions into Eqs.\,(\ref{Maxwell}) and
(\ref{macro}) and equating coefficients at the same exponentials
results in the following eigenvalue problem
 \begin{eqnarray}
&&\hspace{-10mm}\left[
(k+nK)^2-\frac{\varepsilon_b}{c^2}(\omega+n\Omega+m\Omega)^2\right]
{\cal E}_{nm} \nonumber\\
 &&\hspace{22mm}=\frac{4\pi}{c^2}\,(\omega+n\Omega+m\Omega)^2
{\cal P}_{nm}\,,
 \label{Maxwell-Enm}
 \\
&&\hspace{-10mm}\Bigl[\omega_t^2-2i\gamma
(\omega+n\Omega+m\Omega)-(\omega+n\Omega+m\Omega)^2\Bigr]
{\cal P}_{nm}\nonumber\\
&&\hspace{3mm}+2\sigma\omega_t({\cal P}_{n-1,m}+{\cal
P}_{n+1,m})=\frac{\varepsilon_b\,\omega_R^2}{4\pi}\,{\cal
E}_{nm}\,,
 \label{macro-Enm}
 \end{eqnarray}
which has to be solved to find eigenvalues $\omega$ and
eigenvectors ${\cal E}_{nm}$ (and ${\cal P}_{nm}$). It is
convenient however to introducing a new eigenvalue
$\bar{\omega}=\omega+m\Omega$ which makes the eigenvalue problem
$m$-independent, so that the eigenvectors $ {\cal E}_{nm}$ and
${\cal E}_{nm'}$ calculated for different $m$ and $m'$ turn out to
be proportional to each other and corresponding to the same,
$m$-independent eigenvalue $\bar{\omega}$. In other words, all
terms in Eqs.\,(\ref{Enm}) and (\ref{Pnm}) having the same index
$n$ are linearly dependent, which effectively eliminates the
$m$-summation bringing the above series to the form of
Eqs.\,(\ref{En}) and (\ref{Pn}).

Note that the second index $m$ of the formal double Fourier
transform ${\cal E}_{nm}$ and ${\cal P}_{nm}$ has nothing to do
with the index $j$ in $E_{nj}$ and $P_{nj}$ used throughout the
paper to label different partial waves (normalized polariton
eigenstates) while the first index $n$ denote different
Fourier/Bragg components of the same partial wave.

\section{Linearization of the eigenvalue problem}
\label{B}

The eigenvalue problem given by Eqs.\,(\ref{Maxwell-En}) and
(\ref{macro-En}) is nonlinear, both in $\omega$ and $k$, and is
solved either to find the polariton eigenfrequencies in
Sec.\,\ref{sec-eigenfrequencies}, or to find the complex polariton
wave vectors in Sec.\,\ref{sec-quasimomentum}. Let us show that in
both cases this set of equations can be reduced to a standard {\it
linear} eigenvalue problem.

(i) Let us first consider the quasi-energy dispersion $\omega(k)$
in which a real-valued wave vector $k$ is a given number
(parameter). We assume  $\gamma=0$ as done in
Sec.\,\ref{sec-eigenfrequencies}. To linearize the eigenvalue
problem, we introduce two new vectors,
 \begin{eqnarray}
A_n&=& (\omega+\Omega) P_n\,, \\
B_n&=&(\omega+\Omega) (\varepsilon_b E_n  + 4\pi P_n)\nonumber\\
&=&(\omega+\Omega) \varepsilon_b E_n  + 4\pi A_n\,,
 \end{eqnarray}
and substitute them into Eqs.\,(\ref{Maxwell-En}) and
(\ref{macro-En}). The eigenvalue problem then takes the form
 \begin{eqnarray}
\omega P_n&=&-n\Omega P_n+A_n\,, \\
\omega E_n&=& -n\Omega E_n+\frac{B_n}{\varepsilon_b}-\frac{4\pi
A_n}{\varepsilon_b}\,,\\
\omega B_n&=& -n\Omega B_n+c^2(k+n K)^2 E_n\,,  \\
 \omega A_n&=&-n\Omega A_n-\frac{\varepsilon_b \omega_R^2}{4\pi} E_n\nonumber\\
&&+\omega_t^2 P_n +2\sigma\omega_t(P_{n+1}+P_{n-1})\,.
\end{eqnarray}
The above set of equations can be written in a compact matrix
form:
 \begin{equation}
\omega \vec{\mathbb X}=\hat{\mathbb W}\vec{\mathbb X}\,,
 \end{equation}
where $\hat{\mathbb W}$ and $\vec{\mathbb X}$ are, respectively,
$4M\!\times\!4M$ matrix and $4M$-vector defined as (the truncation
of the matrices the and the definition of $M$ are introduced in
Sec.\,\ref{sec-eigenfrequencies})
 \begin{equation}
\hat{\mathbb W}=\left|
\begin{array}{cccc}
-\Omega\hat{n} & 0 & \omega_t^2 \hat{1}+2\sigma \omega_t \hat{f}& 0\\
0 &-\Omega\hat{n} & 0& c^2 (k\hat{1}+K\hat{n})^2 \\
\hat{1}&0 & -\Omega\hat{n}& 0 \\
-4\pi\varepsilon_b^{-1} \hat{1} & \varepsilon_b^{-1} \hat{1}&0& -\Omega \hat{n}\\
\end{array} \right|,
 \label{AppB-matrix}
 \end{equation}
 \begin{equation}
\vec{\mathbb X}=\left|
\begin{array}{c}
\vec{A}\\
\vec{B}\\
\vec{P}\\
\vec{E}\\
\end{array}
\right|\,,
 \label{AppB-vector}
 \end{equation}
$\hat{1}$, $\hat{n}$, and $\hat{f}$ are $M\!\times\!M$ matrices
having the following properties:
 \begin{eqnarray}
 \bigl(\hat{1}\vec{P}\bigr)_n&=&P_n\,, \\
 \bigl(\hat{n}\vec{P}\bigr)_n&=&n P_n\,, \\
\bigl(\hat{f}\vec{P}\bigr)_n&=&P_{n-1}+P_{n+1}\,,
 \end{eqnarray}
and $\vec{A}$, $\vec{B}$, $\vec{P}$, and $\vec{E}$ are
$M$-vectors.

(ii) In the opposite case of the eigenvalue problem to be solved
for the complex-valued momentum $k$, keeping the real frequency
$\omega$ constant, we do essentially the same. However, before
introducing a new vector, we express the polarization $P_n$ from
Eq.\,(\ref{macro-En}) which can be written as
\begin{equation}
  \sum_{m}G_{nm} P_m=E_n\,,
\end{equation}
where
\begin{eqnarray}
G_{nm}&=&\frac{4\pi}{\varepsilon_b\,\omega_R^2}
\biggl\{\Bigl[\omega_t^2-2i\gamma
(\omega+n\Omega)-(\omega+n\Omega)^2\Bigr]\delta_{nm} \nonumber\\
&&+ 2\sigma\omega_t(\delta_{n,m+1}+\delta_{n,m-1}) \biggr\}\,.
 \end{eqnarray}
Noting that $G_{nm}$ does not depend on $k$ (it depends on
$\omega$ only), this matrix can be inverted that results in the
following equation
\begin{equation}
\frac{4\pi}{c^2}(\omega+n\Omega)^2P_n=\sum_m H_{nm} E_m\,,
 \label{appB-H}
\end{equation}
where the matrix $\hat{H}$ has the form
\begin{equation}
H_{nm}=\frac{4\pi}{c^2}(\omega+n\Omega)^2 (G^{-1})_{nm}\,.
\end{equation}
Comparing Eqs.\,(\ref{Maxwell-En}) and (\ref{appB-H}) and also
introducing a new vector ${\vec F}$ defined as
\begin{equation}
F_{n}=\left(k+nK+\sqrt{\varepsilon_b}\,\frac{\omega+n\Omega}{c}\right)E_n\,,
\end{equation}
a linearized matrix eigenvalue problem takes the form
 \begin{equation}
\hat{\mathbb V}\vec{\mathbb Y} =k \vec{\mathbb Y},
 \end{equation}
where $\hat{\mathbb V}$ and $\vec{\mathbb Y}$ are, respectively,
$2M\!\times\!2M$ matrix and $2M$-vector, defined as
\begin{equation}
\hat{\mathbb V}=\left|
\begin{array}{cc} \displaystyle
-\hat{n}K-\sqrt{\varepsilon_b}\,\frac{\omega\hat{1}+\hat{n}\Omega}{c}
& \hat{1}\\
\hat{H} &\displaystyle
-\hat{n}K+\sqrt{\varepsilon_b}\,\frac{\omega\hat{1}+\hat{n}\Omega}{c}\\
\end{array} \right|,
 \label{appB-V}
\end{equation}
\begin{equation}
\vec{\mathbb Y}=\left|
\begin{array}{c}
\vec{E}\\
\vec{F}\\
\end{array}
\right|\,.
 \label{appB-Y}
\end{equation}

The procedure described in Sec.\,\ref{sec-quasimomentum} to refine
the polariton wave vectors $\tilde{\kappa}_{j\neq0}$ using the
best calculated central root $\varkappa(\omega)$, utilizes the
translational property of the modified polariton dispersion
Eq.\,(\ref{k-transl}) and the following linear interpolation
between neighboring points in frequency, $\omega_i\leqslant
\omega-j\Omega\leqslant\omega_{i+1}$:
 \begin{eqnarray}
\tilde{\kappa}_j(\omega) &=& jK+\varkappa(\omega - j
\Omega) \label{interp}\\
&=&\varkappa(\omega_i)+\frac{\varkappa(\omega_{i+1})-\varkappa(\omega_i)}{\omega_{i+1}-\omega_i}
(\omega-j\Omega-\omega_i)\,.\nonumber
 \end{eqnarray}

\section{Analytic evaluation of the band gaps}
\label{C}

When calculating the band gap widths $\Delta_N$ we have to take
into account the effect of mixing (and anticrossing) of two
polariton waves. One wave has the wave vector $k=k_N=NK/2$ and
refers to the $n=0$ replica, the other has the wave vector
$k=k_N-NK=-NK/2$ and refers to $n=-N$ component of the electric
field. To take them into account in the lowest ($N$-th) order we
truncate the tridiagonal matrix equation Eq.\,(\ref{tri}) to
\begin{equation}
\left|
 \begin{array}{ccccc}
 D_0 & -\sigma &0 & \dots  & 0 \\
 -\sigma & D_{-1} & -\sigma & \dots  & 0 \\
 0 & -\sigma & D_{-2} &  \dots  & 0 \\
\vdots& \vdots & \vdots& \ddots & \vdots\\
 0 & 0 & 0 &  \dots & D_{-N} \\
 \end{array}
\right| \cdot\left|
 \begin{array}{c}
 \alpha_0 E_0\\
\alpha_{-1} E_{-1}\\
\alpha_{-2} E_{-2}\\
\vdots\\
\alpha_{-N} E_{-N}\\
 \end{array}
\right| = 0\,,
 \label{appC-trunc}
\end{equation}
where $D_n\equiv D_n(k_N,\omega)$ and
\begin{equation}
D_n(k,\omega)=\frac{1}{2\omega_t}\left[\omega_t^2- \omega^2-
\frac{\omega_R^2}{(k/q)^2-1)}\right] \,,
\end{equation}
neglecting $\gamma$ and $\Omega$ [cf. with Eq.\,(\ref{Dn})], and
$q=\sqrt{\varepsilon_b}\,\omega/c$. We then solve
Eq.\,(\ref{appC-trunc}) and find small deviations of $\omega$ from
the bare gap positions $\omega_N$ given by the equation
$D_0(k_N,\omega_N)=0$. First of all, we find deviations of $D_0$
from zero by solving
\begin{equation}
{\rm det} \left|
 \begin{array}{ccccccc}
 D_0 & -\sigma &0 & \dots & 0 & 0 & 0 \\
 -\sigma & D_{-1} & -\sigma & \dots & 0 & 0 & 0 \\
 0 & -\sigma & D_{-2} &  \dots & 0 & 0 & 0 \\
\vdots& \vdots & \vdots & \ddots &\vdots & \vdots & \vdots\\
 0 & 0 & 0 & \dots & D_{-2} & -\sigma & 0\\
 0 & 0 & 0 &  \dots & -\sigma & D_{-1} & -\sigma \\
 0 & 0 & 0 &  \dots & 0 & -\sigma & D_{0} \\
 \end{array}
\right|=0\,,
 \label{appC-det}
\end{equation}
where the symmetry property $D_s=D_{N-s}$ is used. This results in
the following quadratic equation for $D_0$:
\begin{equation}
D_0^2 S_{11}-2D_0\sigma^2 S_{12}+\sigma^4 S_{22}=0\,,
\end{equation}
where (for the fixed $N$)
\begin{equation}
S_{nm}= {\rm det} \left|
 \begin{array}{cccccc}
 D_{-n} & -\sigma &0 & \dots & 0 & 0 \\
 -\sigma & D_{-n-1} & -\sigma & \dots & 0 & 0 \\
 0 & -\sigma & D_{-n-2} &  \dots & 0 & 0 \\
\vdots& \vdots & \vdots& \ddots &\vdots& \vdots\\
 0 & 0 & 0 &  \dots & D_{-m-1} & -\sigma \\
 0 & 0 & 0 &  \dots & -\sigma & D_{-m} \\
 \end{array}
\right|.
 \label{appC-S}
\end{equation}
Noting that
\begin{equation}
S_{12}^2-S_{11}S_{22}=\sigma^4(S_{23}^2-S_{22}S_{33})=\dots=\sigma^{2(N-2)}\,,
\end{equation}
we find
\begin{equation}
D_0(\omega_N^\pm)=\frac{\sigma^2 S_{12}\pm\sigma^N}{S_{11}}\,,
\end{equation}
where the two values $\omega_N^\pm$ are the energy positions of
the top and the bottom of the $N$-th gap. The gap width
$\Delta_N=|\omega_N^+-\omega_N^-|$ is then proportional to
\begin{equation}
D_0(\omega_N^+)-D_0(\omega_N^-)=\frac{2\sigma^N}{S_{11}}\approx\frac{2\sigma^N}{D^2_{-1}
D^2_{-2}\dots D^s_{-[\frac{N}{2}]}}
\end{equation}
where we have neglected all orders higher than $N$. In the last
equation, $s=1$ and $[\frac{N}{2}]= \frac{N}{2}$ for even $N$, and
$s=2$ and $[\frac{N}{2}]=\frac{N-1}{2}$ for odd $N$. Finally, the
band gap $\Delta_N$ is evaluated using the linear approximation
\begin{equation}
D_0(\omega_N^+)-D_0(\omega_N^-)\approx
(\omega_N^+-\omega_N^-)\left.\frac{\partial D_0}{\partial
\omega}\right|_{\omega=\omega_N}
\end{equation}
where the derivative is calculated and simplified to
\begin{equation}
\frac{\partial D_0}{\partial
\omega}=-\,\frac{\omega_t^2\omega_R^2+(\omega_t^2-\omega^2)^2}{\omega_t\omega\omega_R^2}\,,
\end{equation}
using the condition $D_0(k_n,\omega)=0$ for $\omega=\omega_N$,
taken on the bare polariton dispersion.
\section{Bragg replicas in the limit of small perturbations}
 \label{D}
Substitution of the plane wave expansion Eqs.\,(\ref{En2}) and
(\ref{Pn2}) into the Maxwell and material Equation (\ref{Maxwell})
and (\ref{macro}) results in the following tridiagonal matrix
equation for the electric field components
\begin{equation}
\alpha_{nj} D_{nj} E_{nj}-\sigma
\Bigl(\alpha_{n+1,j}E_{n+1,j}+\alpha_{n-1,j}E_{n-1,j}\Bigr) = 0\,,
 \label{tri2}
\end{equation}
where
 \begin{eqnarray}
 &&D_{nj}=D_n(\tilde{\kappa}_j,\omega)\,,\\
 &&\alpha_{nj}=\alpha_n(\tilde{\kappa}_j,\omega)\,,
 \end{eqnarray}
and $D_n$ and $\alpha_n$ are given by Eqs.\,(\ref{Dn}) and
(\ref{alpha}), respectively. Equation (\ref{tri2}) is similar to
Eq.\,(\ref{tri}) but now includes explicitly the index $j$ which
refers to a specific partial polariton wave. Using the translation
property Eq.\,(\ref{k-transl}) we obtain
 \begin{eqnarray}
D_{n,j+s}(\omega)&=& D_{n+s,j}(\omega-s\Omega)\,,
\\
 \alpha_{n,j+s}(\omega)&=& \alpha_{n+s,j}(\omega-s\Omega)\,,
 \end{eqnarray}
and then, neglecting the acoustic frequency $\Omega$, we are able
to introduce universal vectors $\tilde{E}_n$, $\tilde{D}_n$,
$\tilde{\alpha}_n$, and $\tilde{\kappa}_n$:
 \begin{eqnarray}
\tilde{E}_{n+j}(\omega)&=&E_{nj}(\omega)\,, \\
\tilde{D}_{n+j}(\omega)&=&D_{nj}(\omega)\,, \\
\tilde{\alpha}_{n+j}(\omega)&=&\alpha_{nj}(\omega)\,, \\
\tilde{\kappa}_{j}(\omega)&=&\varkappa(\omega)+jK\,,
 \end{eqnarray}
where $\varkappa(\omega)$ is the central root introduced in
Eq.\,(\ref{kappa}). Then Eq.\,(\ref{tri2}) takes the form
\begin{equation}
\tilde{\alpha}_n \tilde{D}_n \tilde{E}_n-\sigma
\Bigl(\tilde{\alpha}_{n+1}\tilde{E}_{n+1}+\tilde{\alpha}_{n-1}\tilde{E}_{n-1}\Bigr)
= 0
 \label{tri3}
\end{equation}
which is formally equivalent to Eq.\,(\ref{tri}) but has a
different meaning. This is because it effectively couples
components of different partial waves (having different $j$). In
the following we drop all tildes in the above quantities for
simplicity of notations.

To solve this equation perturbatively we note that
$E_{n}=O(\sigma^{|n|})$. Neglecting higher orders in $\sigma$, we
obtain from Eq.\,(\ref{tri3})
 \begin{eqnarray}
&&\alpha_1 D_1 E_1=\sigma\alpha_0E_0\,,
\nonumber \\
&& \alpha_2 D_2 E_2=\sigma\alpha_1 E_1\,, \nonumber\\
 &&\dots \nonumber \\
&&E_{\pm N}=\sigma \frac{\alpha_{\pm (N-1)}}{\alpha_{\pm N} D_{\pm
N}} E_{\pm(N-1)}\,,
 \end{eqnarray}
or, more explicitly,
\begin{equation}
E_{\pm N}=\sigma^N  \frac{\alpha_0}{\alpha_{\pm N}}
\prod_{s=1}^N\frac{1}{D_{\pm s}}\,,
\end{equation}
where we have used the starting value $E_0=1$.

To calculate the reflection, we use the boundary conditions
Eqs.\,(\ref{bc1}) and (\ref{bc2}) which take the form (also
replacing $A_j\to A_{-j}$):
\begin{eqnarray}
\delta_{n,0}+r_n &=& \sum_j A_{-j} E_{n+j}\,,
\label{appD-bc1}\\
2\delta_{n,0} &=& \sum_j A_{-j} E_{n+j}(1+\beta_{n+j})\,,
\label{appD-bc2}
\end{eqnarray}
where
\begin{equation}
\beta_n(\omega)=\frac{\varkappa(\omega)+nK}{q(\omega)}\,,\ \ \ \
q(\omega)=\frac{\omega}{c}\,.
\end{equation}
In the perturbation approach, $A_j$ and $r_n$ have the same
properties as $E_n$, namely $A_{j}=O(\sigma^{|j|})$ and
$r_{n}=O(\sigma^{|n|})$. We calculate them in a similar way,
keeping only the leading-order terms:
 \begin{eqnarray}
A_0&=&\frac{2}{1+\beta_0}\,, \nonumber \\
A_1E_0(1+\beta_0)&=&-A_0 E_1 (1+\beta_1)\,, \nonumber\\
A_2E_0(1+\beta_0)&=&-A_0 E_2 (1+\beta_2)-A_1 E_1 (1+\beta_1)\,,\nonumber\\
&&\dots \nonumber\\
A_{\pm N} &=& -\sum_{s=1}^N A_{\pm(N-s)} E_{\pm
s}\,\frac{1+\beta_{\pm s}}{1+\beta_0}\,,
 \label{appD-A}
 \end{eqnarray}
as follows from Eq.\,(\ref{appD-bc2}), and
 \begin{eqnarray}
r_0&=&A_0 E_0-1= \frac{1-\beta_0}{1+\beta_0}\,, \nonumber \\
 & & \dots  \nonumber \\
 r_{\pm N}&=&\sum_{s=0}^N A_{\pm (N-s)} E_{\pm s}=\sum_{s=1}^N A_{\pm (N-s)} E_{\pm
 s}+ A_N \,,\nonumber \\
 &=& \sum_{s=1}^N A_{\pm (N-s)} E_{\pm s} \frac{\beta_0-\beta_{\pm
 s}}{1+\beta_0}\,,
 \end{eqnarray}
as follows from Eqs.\,(\ref{appD-bc1}) and (\ref{appD-A}).

\section{Analytic evaluation of the decay (interaction) length}
 \label{E}
To calculate the interaction length at the energy of the first
acoustically-induced band gap $\omega=\omega_1$, we take into
account, in the resonant approximation, the contribution of only
two co- and contra-propagating polariton waves. Keeping in
Eq.\,(\ref{En}) only $n=0$ and $n=-1$ terms,
\begin{equation}
E(z,t)=e^{-i\omega t}\Bigl[E_0 e^{i k z}+E_{-1} e^{i(k-K)
z}\Bigr]\,,
\end{equation}
we evaluate the imaginary part $Q$ of the polariton wave vector
\begin{equation}
k=\frac{K}{2}+iQ\,,
 \label{appE-k}
\end{equation}
again, neglecting $\Omega$ and $\gamma$. Then Eq.\,(\ref{tri})
yields
 \begin{eqnarray}
&&\alpha_0 D_0 E_0-\sigma\alpha_{-1} E_{-1} = 0\,,
\nonumber\\
&&\alpha_{-1} D_{-1} E_{-1}-\sigma\alpha_0 E_0 = 0\,,
 \end{eqnarray}
from which we obtain
\begin{equation}
D_0 D_{-1}=\sigma^2\,.
 \label{appE-D}
\end{equation}
Using the property
$D_{-1}(k,\omega)=D_{0}(k^\ast,\omega)=D_{0}^\ast(k,\omega)$ valid
for the $k$ given by Eq.\,(\ref{appE-k}) and introducing the real
and imaginary parts of $D_0$ as $D_0=D_0'+iD_0''$,
Eq.\,(\ref{appE-D}) transforms to
\begin{equation}
D_0'^2+D_0''^2=\sigma\,.
\end{equation}
Since $D_0'=0$ at the gap position $\omega=\omega_1$, the above
equation results in
\begin{equation}
D_0''(k,\omega_1)=\sigma^2\,,
\end{equation}
or in a more explicit equation for $Q$:
\begin{equation}
{\rm
Im}\,\frac{\omega_R^2}{\displaystyle\left(\frac{K/2+iQ}{q}\right)^2-1}=2\omega_t\sigma\,,
\end{equation}
where $q=\sqrt{\varepsilon_b}\omega/c$. Finally, assuming $Q\ll
K$ leads to
\begin{equation}
Q=\frac{2\sigma\omega_t}{\omega_R^2}\,\frac{[(K/2q)^2-1]^2}{K/q^2}=\frac{2\sigma\omega_t\omega_R^2
q^2}{(\omega^2_t-\omega^2)^2}\,.
\end{equation}
The interaction length is then evaluated from the decay of the
electric field intensity and has the form
\begin{equation}
l_{\rm int}=\frac{1}{2Q}\,.
\end{equation}


\begin{thebibliography}{99}
\bibitem{footnote1} This simmetry property fulfils approximately provided that
$\Omega\ll\omega$ and becomes strict assuming $v_s=0$. More
generally, there is the following strict property: if $\omega(k)$
is an eigenvalue of Eqs.\,(\ref{Maxwell-En}) and (\ref{macro-En}),
then $-\omega^\ast(-k^\ast)$ is also an eigenvalue.
\end{thebibliography}
\end{document}